\documentclass[%
 reprint,
 amsmath,amssymb,
 aps,
 prl, 
 noeprint
]{revtex4-2}

\usepackage{graphicx}
\usepackage{dcolumn}
\usepackage{bm}

\usepackage{color}

\def\hn{\hat{\boldsymbol{n}}}
\def\bk{\boldsymbol{k}}
\def\bq{\boldsymbol{q}}
\def\bp{\boldsymbol{p}}
\def\bx{\boldsymbol{x}}
\def\hk{\hat{\boldsymbol{k}}}
\def\hq{\hat{\boldsymbol{q}}}
\def\rd{{\mathrm d}}

\def\TF{{\mathcal T}} 
\def\FFK{{\mathcal K}} 

\begin{document}

\title{Universality of Primordial Anisotropies in Gravitational Wave Background}

\author{Boyuan Jiang$^{1}$}%
\email{rsaito@yamaguchi-u.ac.jp}
\author{Ryo Saito$^{2,3}$}%
\email{rsaito@yamaguchi-u.ac.jp}
\author{Ying-li Zhang$^{1,4,5,3,6}$}
\email{rsaito@yamaguchi-u.ac.jp}
\affiliation{$^{1}$
 School of Physics Science and Engineering, Tongji University, Shanghai 200092, China
}%
\affiliation{$^{2}$
Graduate School of Science and Engineering, Yamaguchi University, Yamaguchi 753-8512,
Japan
}%
\affiliation{$^{3}$
Kavli Institute for the Physics and Mathematics of the Universe (WPI),
UTIAS, The University of Tokyo, Chiba 277-8583, Japan
}%
\affiliation{$^{4}$
Institute for Advanced Study of Tongji University, Shanghai 200092, China
}%
\affiliation{$^{5}$
Asia Pacific Center for Theoretical Physics, Pohang 37673, Korea
}%
\affiliation{$^{6}$
Center for Gravitation and Cosmology, Yangzhou University, Yangzhou 225009, China
}%

\date{\today}

\begin{abstract}

We investigate the general structure of primordial anisotropies in the stochastic gravitational-wave background (SGWB). 
Many mechanisms generating primordial SGWB anisotropies, including fossil and propagation effects, exhibit similar multipole dependences despite their distinct physical origins, implying the presence of a common underlying structure. 
To characterize this structure, we introduce a model-independent framework, the Cosmological Form Factor (CFF), that describes how long-wavelength perturbations modulate short-scale SGWB statistics. 
Within this broad class of mechanisms, we show that the allowed angular and polarization structures of SGWB anisotropies are universally determined by general principles, including statistical isotropy, locality, and the basic properties of long-wavelength perturbations.
In contrast to these universal structures, observables such as amplitudes and frequency dependence retain information about the origin. 
Our results provide a unified and effective description of primordial SGWB anisotropies, clarifying which aspects are universal and which are model dependent, and therefore offer a theoretical foundation for interpreting future gravitational-wave observations. 

\end{abstract}

\maketitle

{\it Introduction and Summary} ---
The detection and characterization of the stochastic gravitational-wave background (SGWB) represents a promising new frontier in cosmology~\cite{Maggiore:2007ulw, Caprini:2018mtu, Christensen:2018iqi}. 
Recent pulsar timing array observations have reported evidence for a nanohertz SGWB signal~\cite{NANOGrav:2023gor, antoniadis2023second, Zic_2023, xu2023searching, Hobbs_2010} and motivated studies of its possible cosmological interpretations~\cite{NANOGrav:2023hvm}, while current and future interferometers will probe complementary frequency ranges~\cite{2015, Acernese_2014, 2019, auclair2023cosmology, ruan2020taiji, li2025gravitational, kawamura2011japanese, KAGRA:2025swt}. 
Distinct primordial sources --- including primordial gravitational waves (PGWs) \cite{Grishchuk:1974ny, Starobinsky:1979ty, Rubakov:1982df} and scalar-induced GWs (SIGWs)~\cite{Ananda:2006af, Baumann:2007zm, Domenech:2021ztg, Assadullahi:2009jc, Kohri:2018awv, Cai:2018dig, Inomata:2018epa, Byrnes:2018txb, Saito:2008jc, Saito:2009jt, Bugaev:2009zh, Garcia-Bellido:2017aan, Bartolo:2018rku} --- and astrophysical sources \cite{Regimbau:2011rp} may produce similar spectra over a finite observational frequency band. 
This motivates exploiting the additional information encoded in its angular correlations~\cite{Romano:2016dpx, LISACosmologyWorkingGroup:2022kbp, Cui:2023dlo}.

\begin{figure*}[!t]
    \centering
    \includegraphics[width=0.9\textwidth]{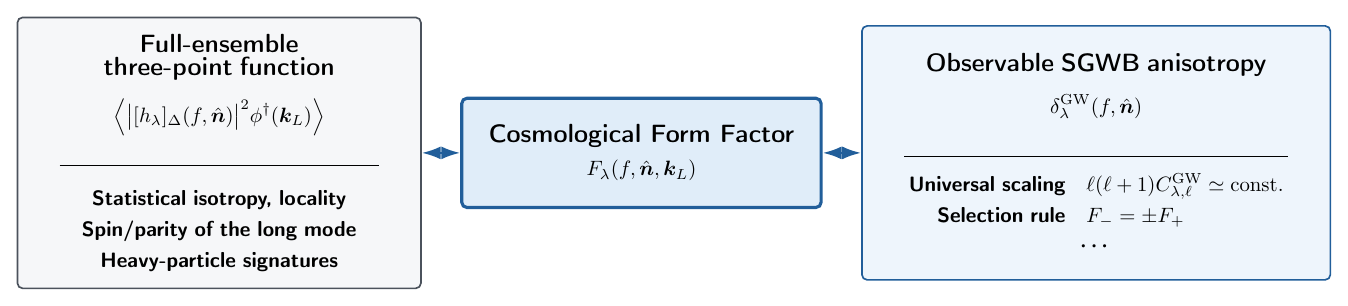}
    \caption{
    Conceptual role of the CFF. 
    The CFF provides a bridge between the full-ensemble long--short--short three-point function and the observable SGWB anisotropy, translating information encoded in the correlator into observable structures. General constraints on the correlator lead, for example, to universal scaling and selection rules. 
    }
    \label{fig:CFF_bridge}
\end{figure*}

The potential importance of large-angle correlations is familiar from the cosmic microwave background (CMB). 
Multipoles $\ell \lesssim 100$ probe scales larger than the horizon at recombination, where the Sachs--Wolfe effect maps primordial superhorizon perturbations onto large-angle temperature anisotropies. 
For a nearly scale-invariant primordial spectrum, this produces the Sachs--Wolfe plateau,
$\ell(\ell+1)C_\ell^{TT}\simeq{\rm const.}$ 
Because this structure is comparatively insensitive to the detailed microphysics of the photon--baryon plasma, it provides a direct probe of primordial correlations. 
Large-angle correlations in the SGWB may similarly provide a clean window into superhorizon physics and, in particular, inflation.

Large-angle SGWB anisotropies have been studied in several settings~\cite{Schulze:2023ich, ValbusaDallArmi:2024hwm}. 
Known mechanisms include fossil effects generated by primordial long--short mode couplings~\cite{Bartolo:2019zvb, Dimastrogiovanni:2021mfs, Dimastrogiovanni:2022afr, Yu:2023jrs, Li:2023qua, Li:2025met, Akama:2024vgu, Li:2023xtl}, 
as well as propagation effects analogous to the Sachs--Wolfe and integrated Sachs--Wolfe effects in the CMB~\cite{Contaldi:2016koz, Bartolo:2019oiq, Bartolo:2019yeu, Li:2023xtl}. 
Remarkably, in a broad class of examples, these physically distinct mechanisms yield the same Sachs--Wolfe-like plateau,
$\ell(\ell +1)C^{\rm GW}_{\ell} \simeq {\rm const}$. 
This raises the question of whether a universal structure underlies these apparently different phenomena.

To identify what is universal, one must first clarify which angular quantity is actually observable and what produces its direction dependence. 
In this Letter, 
we show how these questions lead to a model-independent description of large-angle SGWB anisotropies, which we call the {\bf Cosmological Form Factor (CFF)}. 
Phase-sensitive correlations of GW amplitudes are generally washed out by the coarse graining inherent in finite temporal, frequency, and angular resolution~\cite{Allen:1999xw, Bartolo:2018evs, Bartolo:2018rku, Margalit:2020sxp, Wu:2022kld}. 
The relevant angular information must therefore be extracted from a phase-insensitive quantity, namely the GW intensity. 
While CMB observations effectively provide intensity and polarization maps, GW experiments measure amplitude-level time series, such as interferometric strain or pulsar timing residuals, from which an intensity map must be statistically constructed. 
We define a local intensity estimator by using GWs within a given angular and frequency resolution element. 
This construction averages over unresolved short-wavelength modes while keeping the long-wavelength mode realization fixed. 
The directional variations that remain in this local average constitute the observable anisotropies studied in this work. 

The distinction between this local average and the full ensemble average is essential. 
Within the standard inflationary settings, the full ensemble average is isotropic. 
A fixed realization of a long-wavelength mode, however, modulates the local statistics of the short-scale GWs and thereby biases the locally averaged intensity. 
The conditional long-mode bias of the local intensity is therefore the physical origin of the observable anisotropies.

We formulate this conditional bias as a response problem and parameterize it by the CFF. 
Analogous to the form factors in scattering theory~\cite{Perdrisat:2006hj} and the tidal Love numbers in gravitational physics~\cite{Chakraborty:2026qru, Rodriguez:2026iot}, 
the CFF characterizes how long modes, acting as external fields, probe the SGWB through its observable response: 
it encodes physics in an observable without specifying its detailed microscopic origin. 

As summarized in Fig.~\ref{fig:CFF_bridge}, the CFF provides a bridge between the observed anisotropy and correlation functions defined in the full ensemble. 
It directly determines the observable anisotropy, while being related to a long--short--short three-point function defined through the full ensemble average. 
General constraints that are most naturally imposed on this correlator---including statistical homogeneity and isotropy, locality, and the spin and parity of the long mode---are thereby translated into constraints on the angular and polarization structures of the SGWB anisotropies. 
For nearly scale-invariant scalar long modes, this chain of constraints yields the Sachs--Wolfe-like plateau, whereas the amplitude and frequency dependence retain information about the underlying process. 
The CFF thus separates universal structures dictated by general principles from process-dependent information. 

\medskip

\noindent
{\it Angular Features of SGWB} ---
Following Ref.~\cite{Romano:2016dpx}, we describe the anisotropic features of the SGWB by the direction-dependent power spectrum, 
\begin{align}\label{eq:aPh}
&\left\langle
h_{\lambda}^{\dagger}(f,\hn)
h_{\lambda'}(f',\hn')
\right\rangle
=  \notag \\
&\qquad \qquad \frac{\mathcal P_{\lambda}(f,\hn)}{2}\,
\delta_{\lambda\lambda'}\,
\delta(f-f')\,
\delta^{(2)}(\hn,\hn') \,.
\end{align}
allowing for the helicity dependence~\footnote{
For simplicity, we retain only the helicity-diagonal components
in the Letter. The full polarization matrix, including its off-diagonal components, is discussed in the Supplemental Material.}.
Here, $f$ is the observed GW frequency, $\hn$ denotes the
sky direction from which the GW arrives, and $\lambda=\pm$
labels the circular polarization~\footnote{
Our frequency--direction and polarization conventions follow the standard SGWB convention of Ref.~\cite{Romano:2016dpx}. 
The convention is summarized in the Supplemental Material.}.
Taking the isotropic background to be unpolarized, we decompose
\begin{align}
\mathcal P_{\lambda}(f,\hn)
=
\frac{S_h(f)}{8\pi}
\left[
1+\delta^\text{GW}_{\lambda}(f,\hn)
\right] \,,
\end{align}
where
$S_h(f) \equiv \sum_\lambda \int \rd^2\hn\, \mathcal P_{\lambda}(f,\hn)$ is the standard one-sided strain power spectral density.
Thus, $\delta^{\rm GW}_{\lambda}(f,\hn)$ represents the
fractional anisotropy of the helicity-resolved strain power. 
At fixed frequency, strain power, spectral intensity, and energy density differ only by direction-independent factors and therefore have the same fractional anisotropies. 
The relation to the spatial Fourier modes and the definitions in other literature is given in the Supplemental Material.

The statistical anisotropies $\delta^\text{GW}_\lambda(f,\hn)$ can arise from several mechanisms, including:

\begin{itemize}

    \item {\bf Propagation effects} \cite{Contaldi:2016koz, Bartolo:2019oiq, Bartolo:2019yeu, Li:2023xtl}: 
    
    Gravitational waves experience anisotropic distortions as they propagate through the inhomogeneous universe, analogous to Sachs--Wolfe (SW) and integrated Sachs--Wolfe (ISW) effects in the CMB.

    \item {\bf Fossil effects (non-Gaussian effects)} \cite{Li:2023xtl, Bartolo:2019zvb, Dimastrogiovanni:2021mfs, Dimastrogiovanni:2022afr, Yu:2023jrs, Li:2023qua, Akama:2024vgu, Li:2025met}: 
    
    Mode couplings to cosmological-scale perturbations can imprint directional dependence in the initial SGWB statistics. This effect is especially pronounced in the presence of primordial non-Gaussianity, which couples modes of different scales.
    
\end{itemize}

\noindent

The SGWB anisotropies at each frequency can be characterized
by the angular power spectrum,
\begin{align}\label{eq:Clgw}
\left\langle
[\delta^{\rm GW}_\lambda(f)]^\dagger_{\ell m}
[\delta^{\rm GW}_\lambda(f)]_{\ell' m'}
\right\rangle
=
\delta_{\ell\ell'}\delta_{mm'}
C^{\rm GW}_{\lambda,\ell}(f) \,,
\end{align}
where $[\delta^{\rm GW}_\lambda(f)]_{\ell m}$ denotes the
spherical-harmonic coefficients of
$\delta^{\rm GW}_\lambda(f,\hn)$ with respect to $\hn$. 
Analogous to the CMB, the angular power spectrum $C^\text{GW}_{\lambda, \ell}(f)$ quantifies the statistical variance of the anisotropies as a function of angular scale. 
Intriguingly, $C^\text{GW}_{\lambda, \ell}(f)$ commonly exhibits the plateau behavior,
\begin{align}\label{eq:scaling}
\ell(\ell+1) C^\text{GW}_{\lambda, \ell}(f) \simeq {\rm const},
\end{align}
across a broad class of primordial mechanisms. 
This behavior appears irrespective 
of the origins of the gravitational waves (PGWs or SIGWs) or the generation mechanism of the anisotropies (propagation or fossil effects)~\cite{Contaldi:2016koz, Bartolo:2019oiq, Bartolo:2019yeu, Li:2023xtl, Bartolo:2019zvb, Dimastrogiovanni:2021mfs, Dimastrogiovanni:2022afr, Yu:2023jrs, Li:2023qua, Li:2025met}. 
Our new framework, the Cosmological Form Factor (CFF), is designed to isolate such universal structures from observables sensitive to underlying physics.
\medskip

\begin{figure}[t]
    \centering
    \includegraphics[width=\linewidth]{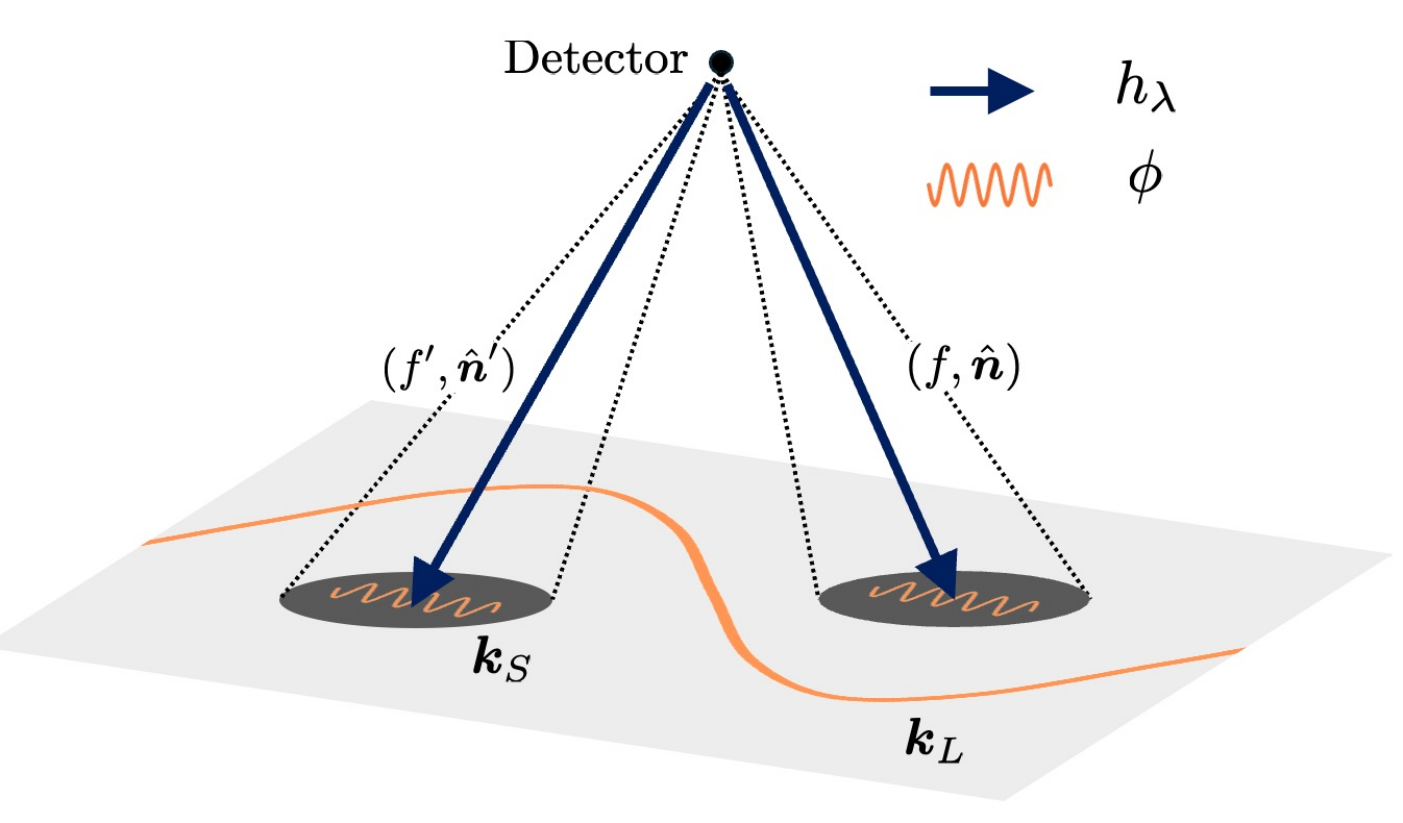}
    \caption{Schematic illustration of the origin of SGWB anisotropies from long-wavelength perturbations. Gravitational waves arriving from sky directions $\hn$ and $\hn'$ probe different local environments due to the presence of a long mode $\phi_L\equiv\phi(\bk_L)$, resulting in statistical anisotropies.
    }
    \label{fig:local_sky}
\end{figure}

\noindent
{\it Averaging and Observable Anisotropy} ---
Equations~\eqref{eq:aPh} and \eqref{eq:Clgw} involve two distinct levels of averaging.  
The anisotropy $\delta^\text{GW}_\lambda$ is defined in Eq.~\eqref{eq:aPh} after an average has been taken, yet its correlations are averaged again in Eq.~\eqref{eq:Clgw}.  
What remains stochastic after the first average, and what is being averaged at each stage?  
In the standard inflationary scenarios, 
interpreting the first bracket as the full ensemble average would eliminate any direction dependence by the statistical isotropy, leaving no anisotropy.  
The first average must therefore have a different operational meaning.  
To identify it, we must return to the observable itself and ask what is actually measured in an SGWB anisotropy measurement. 

An ensemble average is not itself observable, since we have access to only one realization of the SGWB.  
It must instead be inferred from an estimator constructed by combining statistically equivalent GW modes.  
For an anisotropy measurement, modes from all sky directions should not be combined in constructing an estimator: the intensities associated with different directions are precisely the quantities to be compared.  
The observational counterpart of the first average in Eq.~\eqref{eq:aPh} must therefore be a local estimator, constructed only from GWs within a finite angular resolution element and an observed frequency band.

We represent this finite resolution by the filtered amplitude
\begin{align}\label{eq:smoothed h}
&[h_\lambda]_\Delta(f,\hn)
\equiv \notag \\
&\qquad \int \rd f'
\int \rd^2\hn'\,
W_\Delta(f,\hn;f',\hn')\,
h_\lambda(f',\hn') \,,
\end{align}
where $W_\Delta$ is a window function describing the finite
frequency and angular resolutions.
The resulting quadratic quantity provides a local estimator
of the GW intensity,
\begin{align}\label{eq:estimator}
\left|[h_\lambda]_\Delta(f,\hn)\right|^2
\simeq
\left\langle
\left|[h_\lambda]_\Delta(f,\hn)\right|^2
\right\rangle_{\rm local},
\end{align}
where $\simeq$ denotes equality up to the statistical
fluctuations of the estimator. 
A long-wavelength perturbation coherent across that element, however, is not averaged away.  
Instead, it acts as a common background that biases the statistics entering the local intensity estimator.  
Because this bias varies between different sky directions, it constitutes the observable anisotropies. 

This observation provides an operational long--short split.
We denote by $S$ the short sector, which is averaged over by the local estimator \eqref{eq:estimator}.
This sector includes the observed GW modes $h_\lambda$ within the angular and frequency resolution element, as well as any other short-scale degrees of freedom relevant to their statistics.
The anisotropy-generating degree of freedom is a long-wavelength primordial perturbation, which we denote by $L$, whose realization remains fixed during the local average \eqref{eq:estimator}.
In general, this long mode may be scalar, vector, or tensor---and may itself be a long-wavelength GW. 
In this Letter, we restrict our analysis to a scalar long mode $\phi_L$, 
leaving vector and tensor cases for future work. 
The first average $\langle\cdot\rangle_{\rm local}$ is therefore a conditional average over the entire short sector, including $h_\lambda$ and $\phi_S$, at fixed $\phi_L$.
We denote it by $\langle\cdot\rangle_{S,\phi_L}$.
Equation~\eqref{eq:Clgw} then performs the remaining average over the long modes, denoted by $\langle\cdot\rangle_L$.
Determining the SGWB anisotropies thus reduces to determining how $\phi_L$ biases the local intensity statistics entering Eq.~\eqref{eq:aPh}, as inferred from the quadratic estimator in Eq.~\eqref{eq:estimator}.

\medskip

\noindent
{\it Cosmological Form Factor (CFF)} --- 
Our key step is to recast this question as a response problem: we regard the conditional expectation value of the estimator in Eq.~\eqref{eq:estimator} as a functional of the long-mode configuration $\phi_L$.
In the reference configuration $\phi_L=0$, statistical isotropy gives a direction-independent intensity.
Upon turning on a long mode, the locally averaged intensity generally acquires a direction-dependent bias.
The observable anisotropies can therefore be characterized by the response of the local intensity to a variation of the long-mode background.
Expanding this functional around $\phi_L=0$ to linear order, we parameterize the response as
    \begin{align}\label{eq:CFF}
    &\left\langle
    \left|[h_\lambda]_\Delta(f,\hn)\right|^2
    \right\rangle_{S,\phi_L}
    = \nonumber\\
    &\quad
    \frac{S_h(f)\,
    N_\Delta}{16\pi}
    \left[
    1+
    \int\frac{\rd^3\bk_L}{(2\pi)^3}\,
    F_\lambda(f,\hn,\bk_L)\,
    \phi(\bk_L)
    \right] \,,
    \end{align}
where $N_\Delta$ is the normalization factor
    \begin{align}\label{eq:NDelta}
    N_\Delta
    \equiv
    \int \rd f'
    \int \rd^2\hn'\,
    \left|
    W_\Delta(f,\hn;f',\hn')
    \right|^2 \,.
    \end{align}
Comparing Eq.~\eqref{eq:aPh} with Eq.~\eqref{eq:CFF}, the induced SGWB anisotropies are identified as
    \begin{align}\label{eq:delta CFF rel}
    \delta^{\rm GW}_\lambda(f,\hn)
    =
    \int\frac{\rd^3\bk_L}{(2\pi)^3}
    F_\lambda(f,\hn,\bk_L)\phi(\bk_L)\,.
    \end{align}

Importantly, the CFF provides a phenomenological description of observable anisotropies without specifying the detailed production mechanism. 
In this sense, $F_\lambda(f, \hn,\bk_L)$ plays a role analogous to the form factor in scattering theory, encoding microscopic information into observables. 
We therefore refer to it as the Cosmological Form Factor (CFF). 
The response interpretation adopted here parallels long-mode response approaches to large-scale-structure power spectra~\cite{Barreira:2017sqa, Chiang:2017vsq} and clustering fossils~\cite{Jeong:2012df}, as well as fossil-type modulations of SGWB power~\cite{Dimastrogiovanni:2021mfs}. 
Our formulation makes the underlying long-short response structure explicit and shows that the same framework can accommodate any SGWB anisotropies induced by long-wavelength perturbations, including both fossil and propagation effects. 

For a given underlying mechanism, the CFF can be computed explicitly as illustrated by examples presented in the Supplemental Material. 
The central point of this work, however, is that the structure of the CFF is strongly constrained by general principles. 
Multiplying Eq.~\eqref{eq:CFF} by the long-mode field $\phi_L^\dagger$ and performing the remaining long-mode average relates the CFF to the full-ensemble three-point function:
    \begin{align}\label{eq:bispectrum}
        &F_\lambda(f, \hn, \bk_L) = \notag \\
        &\qquad \frac{16\pi}{S_h(f)P_\phi(k_L) N_\Delta}\, \langle |[h_\lambda]_\Delta(f, \hn)|^2 \phi^\dagger(\bk_L) \rangle \,.
    \end{align}
Upon relating $h_\lambda(f,\hn)$ to the spatial Fourier modes, this three-point function is directly connected to the conventional long--short--short bispectrum in Fourier space. 
The precise derivation is given in the Supplemental Material. 

This relation is crucial because it connects the observable anisotropies in Eq.~\eqref{eq:delta CFF rel} to a full-ensemble bispectrum through the CFF.
The advantage of Eq.~\eqref{eq:bispectrum} is that the bispectrum is a standard correlation function whose allowed structure can be directly constrained by general principles, including statistical isotropy and locality.
Through the CFF, these constraints are transferred to the observable SGWB anisotropies.
Hereafter, we investigate these constraints.

\medskip

\noindent

{\it Symmetry Constraint} ---
Statistical isotropy of the underlying bispectrum \eqref{eq:bispectrum} implies that the directional dependence of the CFF can enter only through rotational invariants constructed from $\hn$ and $\bk_L$. 
Its angular dependence must therefore take the form
\begin{align}\label{eq:isotropy}
F_\lambda(f, \hn,\bk_L) = 
F_\lambda(f,k_L,\hn\cdot\hk_L)\, .
\end{align}

Rotational selection rules further constrain the helicity dependence.
For the scalar long mode considered here, the allowed contributions can be decomposed into parity-even and parity-odd structures, satisfying $F_-=F_+$ and $F_-=-F_+$, respectively.
A general CFF may contain both contributions.
A detailed derivation of these selection rules is given in the Supplemental Material.

\medskip

\noindent

{\it Locality Constraint} ---
Locality constrains the squeezed bispectrum in Eq.~\eqref{eq:bispectrum}: at each source time, the short-mode statistics can depend on the scalar long mode only through its value and spatial derivatives evaluated within the corresponding local patch of the past light cone.
The relevant quantities are therefore
\begin{align}
\phi_L(\tau_s, \bx_s),\quad
\nabla\phi_L(\tau_s, \bx_s),\quad
\nabla\nabla\phi_L(\tau_s, \bx_s),\quad \ldots
\end{align}
where $\bx_s$ denotes the position on the local patch of the past light cone corresponding to $\hn$: $\bx_s=\hn(\tau_0-\tau_s)$. 
Here, $\tau_0$ is the present conformal time and $\tau_s$ is the conformal time at which the corresponding anisotropies are generated.
From dimensional considerations, derivative terms are expected to be suppressed by factors of $k_L\tau_s$ or $k_L/(2\pi f)$.
The smallness of $k_L\tau_s$ reflects the fact that the long modes are superhorizon at the source time $\tau_s$.
Compared with the CMB, primordial SGWB observations typically have lower angular resolution and probe earlier source times, so that the long modes relevant to observable anisotropies are well outside the horizon.
This hierarchy ensures the validity of the derivative expansion implied by locality and justifies its truncation at leading order.

Taking the isotropy requirement in Eq.~\eqref{eq:isotropy} into account, the leading-order bias term in Eq.~\eqref{eq:CFF} must take the form
\begin{align}
\int \rd\tau_s\,
\FFK_\lambda(f,\tau_s)
\left.
\phi_L(\tau_s, \bx_s)
\right|_{\bx_s=\hn(\tau_0-\tau_s)} \,,
\end{align}
which leads to
\begin{align}
&F_\lambda(f,k_L,\hn\cdot\hk_L)
= \nonumber \\
&\qquad \int \rd\tau_s\,
\FFK_\lambda(f,\tau_s)
\TF_\phi(k_L,\tau_s)
e^{i(\hn\cdot\hk_L)k_L(\tau_0-\tau_s)} \,,
\end{align}
by using $\phi_L(\tau_s, \bk_L) =\TF_\phi(k_L,\tau_s)\phi(\bk_L)$. 
Here, $\TF_\phi(k_L,\tau_s)$ is the transfer function of the scalar long mode, while $\FFK_\lambda(f,\tau_s)$ is a response kernel at source time $\tau_s$,  encoding the process-dependent physics.

Thus, statistical isotropy and locality leave no process-dependent functional freedom in the $\hn$ dependence of the CFF: all process dependence is confined to the direction-independent response kernel $\FFK_\lambda(f,\tau_s)$.
Through Eq.~\eqref{eq:delta CFF rel}, the angular structure of the SGWB correlations in Eq.~\eqref{eq:Clgw} therefore directly reflects the statistical structure of the long mode $\phi_L$.
The short-scale dynamics determines the response amplitude, frequency dependence, and source-time weighting, but does not generate an independent angular structure.

\medskip

\noindent
{\it Universal Scaling} --- 
With the directional dependence of the CFF uniquely fixed by statistical isotropy and locality, the angular power spectrum in Eq.~\eqref{eq:Clgw} takes the form
    \begin{align}\label{eq:Cl form}
        C^\text{GW}_{\lambda, \ell}(f) = 
        \frac{2}{\pi}\int k_L^2 {\rm d}k_L P_\phi(k_L) |\Delta^\text{GW}_{\lambda, \ell}(f,k_L)|^2 \,,
    \end{align}
where
    \begin{align}
        &\Delta^\text{GW}_{\lambda, \ell}(f,k_L)
        \equiv \notag \\
        &\qquad \int \rd\tau_s~ \FFK_\lambda(f,\tau_s)\TF_\phi(k_L,\tau_s) j_\ell[k_L(\tau_0-\tau_s)] \,,
    \end{align}
and $j_\ell$ denotes the $\ell$-th spherical Bessel function. 
If the scalar long mode is the primordial curvature perturbation, the scales relevant to low-$\ell$ SGWB anisotropies lie within the CMB regime, where its primordial power spectrum is nearly scale invariant: $P_\phi(k_L)\propto k_L^{-3}$. 
For the superhorizon long modes relevant to large-angle anisotropies, $k_L\tau_s \ll 1$, and hence $j_\ell[k_L(\tau_0-\tau_s)] \simeq j_\ell(k_L\tau_0)$.
Moreover, the long-mode transfer function is effectively scale independent, $\TF_\phi(k_L,\tau_s)\simeq \TF_\phi(\tau_s)$. 
Under these conditions, 
the $\ell$-dependence of the angular power spectrum \eqref{eq:Cl form} is determined as
    \begin{align}
        C^\text{GW}_{\lambda, \ell}(f) 
        \propto  \int \frac{{\rm d}k_L}{k_L} [j_\ell(k_L \tau_0)]^2 \propto [\ell(\ell+1)]^{-1} \,,
    \end{align}
which yields the universal behavior \eqref{eq:scaling}. 

\medskip

\noindent
{\it Extensions} ---
While we have focused on a scalar primordial field $\phi$, 
our framework can be naturally extended to higher-spin fields.
Tensor fields, for instance, would generate distinct angular patterns  \cite{Ricciardone:2017kre, Dimastrogiovanni:2018uqy, Dimastrogiovanni:2019bfl}.
The symmetry constraints become even more powerful in these cases, restricting the allowed angular and polarization structures of the CFF.

An especially interesting application is to cosmological collider
physics~\cite{Chen:2009zp,Arkani-Hamed:2015bza,Lee:2016vti}:
the CFF dictionary offers a direct route for translating nonanalytic
squeezed-limit signatures of massive particles into characteristic
frequency, angular, and polarization dependences of SGWB anisotropies,
with a recent massive-spin-two calculation providing an explicit
primordial tensor example~\cite{Bhattacharya:2026spa}.

Our framework can also be systematically extended beyond linear order in $\phi_L$ by introducing an $n$-th-order form factor at each order in the expansion in Eq.~\eqref{eq:CFF}.
These higher-order responses can be conveniently organized diagrammatically, with an $n$-th-order form factor acting as an effective vertex that couples $\delta^\text{GW}_\lambda$ to $n$ long modes.
A systematic development of this diagrammatic formulation will be presented in future work.

\medskip

\noindent
{\it Conclusion and Outlook} ---
Our results clarify both the operational meaning of observable SGWB anisotropies and how the information they contain is organized. 

We identify the observable anisotropies on large scales as the long-mode-induced conditional bias that remains after the local average. This identification also provides a practical criterion for estimator construction: short GW modes should be averaged to obtain a phase-insensitive local intensity, without averaging over the long-mode-induced directional dependence that constitutes the signal. 

We introduced the Cosmological Form Factor (CFF) as a systematic and model-independent description of SGWB anisotropies. 
By treating the locally averaged intensity as a functional of the long-mode background, the CFF encodes this bias as a response. 
The CFF translates this operational picture into templates for statistical inference.
Its symmetry- and locality-constrained angular and polarization structures identify modes that share a common response.
Because only one realization of the Universe is observable, combining such modes increases the effective number of samples and reduces the estimator variance.
At the same time, the process-dependent response amplitude, frequency dependence, and source-time weighting retain the information needed to distinguish the underlying mechanisms.
The first two can be probed from SGWB observations alone, whereas the generation epoch, encoded in the source-time dependence of $\FFK_\lambda(f,\tau_s)$, can be constrained through cross-correlations with external cosmological probes---such as the CMB, galaxy surveys, or weak lensing~\cite{Adshead:2020bji,Malhotra:2020ket,Braglia:2021fxn, Bravo:2025csu}. 

Extensions to higher-spin long modes and higher-order responses will be developed in forthcoming work.

\smallskip

\noindent
{\it Acknowledgements} ---
R.S. is supported by JSPS KAKENHI JP23H01171 and JP23K25868, and Shanghai Sci-tech Co-research Program 25HB2700100. B.J. and Y.Z. are supported by the Fundamental Research Funds for the Central Universities, and by the National Natural Science Foundation of China (NSFC) Grants Nos. 12475060 and W2611007, Project 24ZR1472400 sponsored by Natural Science Foundation of Shanghai, and Shanghai
Pujiang Program 24PJA134.
The authors thank the organizers and participants of the workshop at the Yukawa Institute for Theoretical Physics (YITP-T-26-05) for providing an opportunity to develop these ideas.

\nocite{wang2024reconstructing}
\bibliographystyle{apsrev4-2}
\bibliography{AGWB}

\clearpage
\onecolumngrid
\setcounter{section}{0}
\setcounter{subsection}{0}
\setcounter{secnumdepth}{3}
\setcounter{equation}{0}
\setcounter{figure}{0}
\setcounter{table}{0}
\renewcommand{\thesection}{\Roman{section}}
\renewcommand{\thesubsection}{\Alph{subsection}}
\renewcommand{\theequation}{S\arabic{equation}}
\renewcommand{\thefigure}{S\arabic{figure}}
\renewcommand{\thetable}{S\arabic{table}}

\begin{center}
{\large\bfseries Supplemental Material for\\[0.35em]
\textit{Universality of Primordial Anisotropies in Gravitational Wave Background}\par}
\vspace{0.8em}
Boyuan Jiang,\quad Ryo Saito,\quad and Ying-li Zhang
\end{center}
\vspace{1em}

\section{Conventions and Representations of SGWB Anisotropies}
\label{sec:SM_conventions}

This section summarizes the relation between the frequency--direction representation used in the Letter, the spatial Fourier representation, and other variables commonly used to describe SGWB anisotropies.

\subsection{Frequency--direction and spatial Fourier decompositions}
\label{sec:ffn relations}

In the Letter, we describe the SGWB in terms of the observed frequency $f$ and sky direction $\hn$, following the standard convention used in observational analyses. 
Using $\hn$ to denote the direction from which a GW arrives, the frequency--direction decomposition is
\begin{align}
h_{ij}(t,\bx)
=
\sum_{\lambda=\pm}
\int_{-\infty}^{\infty}\rd f
\int\rd^2\hn\,
h_\lambda(f,\hn)
e^\lambda_{ij}(-\hn)
e^{-2\pi if(t+\hn\cdot\bx)} \,.
\label{eq:frequency_direction_decomposition}
\end{align}
Thus, the corresponding GW propagation direction is $-\hn$.
The positive- and negative-frequency branches are not independent but are related by the reality condition,
\begin{align}\label{eq:reality relation}
h_\lambda^\dagger(f,\hn)
=h_{-\lambda}(-f,\hn) \,.
\end{align}

The spatial Fourier representation, on the other hand, makes the consequences of statistical homogeneity transparent. 
In the spatial Fourier basis, statistical homogeneity is directly encoded by momentum conservation. 
By contrast, the $(f,\hn)$ representation is anchored to a chosen observer, making the same symmetry less manifest. 
In terms of the conformal time $\tau$ and comoving coordinates $\bx$, the tensor perturbation is expanded as
\begin{align}
h_{ij}(\tau,\bx)
=
\sum_{\lambda=\pm}
\int\frac{\rd^3\bq}{(2\pi)^3}\,
h_\lambda(\tau,\bq)
e^\lambda_{ij}(\hq)
e^{i\bq\cdot\bx},
\label{eq:Fourier_decomposition}
\end{align}
where $\bq=q\hq$ is the comoving wave vector.
In the subhorizon regime around the observation epoch, each Fourier mode can be decomposed into positive- and negative-frequency components as
\begin{align}
h_\lambda(\tau,\bq)
=
h^{(p)}_\lambda(\bq)
e^{-iq(\tau-\tau_0)}
+
h^{(n)}_\lambda(\bq)
e^{iq(\tau-\tau_0)} \,.
\label{eq:frequency_branches}
\end{align}
The cosmic and conformal times satisfy $\rd t=a(\tau)\rd\tau$.
Over an observation period much shorter than the Hubble time,
\begin{align}
\tau-\tau_0
\simeq
\frac{t}{a(\tau_0)}
=t \, ,
\end{align}
where we set the beginning of the observation to $t=0$ and adopt $a(\tau_0)=1$.
The comoving wavenumber can therefore be identified with the observed physical angular frequency,
\begin{align}
q=2\pi|f| \,.
\end{align}

Matching Eqs.~\eqref{eq:frequency_direction_decomposition}--\eqref{eq:frequency_branches} gives~\cite{Wu:2022kld}
\begin{align}
h_\lambda(f,\hn)
=
\begin{cases}
f^2h^{(p)}_\lambda(-2\pi |f|\hn)
& (f>0) \,,\\
f^2h^{(n)}_{-\lambda}(2\pi|f|\hn)
& (f<0) \,.
\end{cases}
\label{eq:Fourier_frequency_dictionary}
\end{align}
The factor $f^2$ originates from the change of integration measure, while the helicity reversal on the negative-frequency branch follows from $e^\lambda_{ij}(-\hn)=e^{-\lambda}_{ij}(\hn)$. 
Accordingly, the positive-frequency component of a Fourier mode with wave vector $\bq$ describes a GW arriving from the sky direction $-\hq$, whereas its negative-frequency component is represented by $f=-q/(2\pi)$ and $\hn=\hq$. 
The reality condition \eqref{eq:reality relation} implies that the negative-frequency amplitudes do not represent independent degrees of freedom at the level of the linear GW field: 
    \begin{align}\label{eq:reality relation pn}
        [h_\lambda^{(p)}]^\dagger(\bq) = h_\lambda^{(n)}(-\bq) \,.
    \end{align}
It should be noted that this does not mean that all quadratic information is contained in the usual normal spectral density $S_h(f)$: correlations between the positive- and negative-frequency components can give rise to correlations between antipodal directions $A_h(f)$~\cite{Wu:2022kld}.
As shown in Ref.~\cite{Wu:2022kld}, these correlations are washed out by finite frequency resolution and angular coarse graining.
We therefore omit the corresponding antipodal term from Eq.~(1) of the Letter and Eq.~\eqref{eq:aPh SM} below.

\subsection{Dictionary of conventions for SGWB anisotropies}

We characterized the SGWB anisotropies through the direction-dependent power spectrum [Eq.~(1) in the Letter]: 
    \begin{align}\label{eq:aPh SM}
        \langle
        h_{\lambda}^{\dagger}(f,\hn)
        h_{\lambda'}(f',\hn')
        \rangle
        =  
        {\mathcal P}_{\lambda}^{(D)}(f,\hn)\,
        \delta_{\lambda\lambda'}\,
        \delta(f-f')\,
        \delta^{(2)}(\hn,\hn') \,,
    \end{align}
where we have defined the power spectrum ${\mathcal P}_{\lambda}^{(D)}(f,\hn)$ as a double-sided quantity. 
The corresponding single-sided spectral
density is defined by
\begin{align}\label{eq:one-sided-definition}
\mathcal P_{\lambda}(f,\hn)
\equiv
\mathcal P_{\lambda}^{\mathrm{(D)}}(f,\hn)
+
\mathcal P_{-\lambda}^{\mathrm{(D)}}(-f,\hn) = 2\mathcal P_{\lambda}^{\mathrm{(D)}}(f,\hn)
\qquad (f>0) \,,
\end{align}
where the last equality follows from the reality condition \eqref{eq:reality relation}. 
Restricting to $f>0$, 
Eq.~\eqref{eq:aPh SM} can be written in terms of the single-sided spectral density as
    \begin{align}\label{eq:aPh SM one-sided}
        \langle
        h_{\lambda}^{\dagger}(f,\hn)
        h_{\lambda'}(f',\hn')
        \rangle
        =  
        \frac{{\mathcal P}_{\lambda}(f,\hn)}{2}\,
        \delta_{\lambda\lambda'}\,
        \delta(f-f')\,
        \delta^{(2)}(\hn,\hn') \qquad (f>0) \,.
    \end{align}
This expression is used in the Letter. 
We define the fractional anisotropy $\delta^\text{GW}_{\lambda}(f,\hn)$ by
    \begin{align}\label{eq:delta GW}
        \mathcal P_{\lambda}(f,\hn)
        \equiv 
        \frac{S_h(f)}{8\pi}
        \left[
        1+\delta^\text{GW}_{\lambda}(f,\hn)
        \right] \,.
    \end{align}
Here, we allow for the helicity dependence, while Ref.~\cite{Romano:2016dpx} assumes equal power in the two helicities: $\mathcal P_{+}= \mathcal P_{-} = \mathcal P/2$. 

As we explained in the Letter, the average in Eq.~\eqref{eq:aPh SM one-sided} should be understood as the local average $\langle \cdot \rangle_{S,\phi_L}$ and the observable amplitude is represented by the filtered amplitude,
    \begin{align}\label{eq:filtered amplitude}
        [h_\lambda]_\Delta(f,\hn)
        \equiv
        \int \rd f'\int\rd^2\hn'\,
        W_\Delta(f,\hn;f',\hn')h_\lambda(f',\hn') \,.
    \end{align}
Then, the main observable is given by the following helicity-resolved intensity, 
    \begin{align}\label{eq:intensity SM}
        \langle
        \left|[h_\lambda]_\Delta(f,\hn)\right|^2
        \rangle_{S,\phi_L}
        =
        \frac{S_h(f)N_\Delta}{16\pi}
        \left[
        1+\delta_\lambda^{\rm GW}(f,\hn)
        \right] \qquad (f>0) \,.
    \end{align}
The delta functions in Eq.~\eqref{eq:aPh SM} represent the ideal infinite-resolution limit; in an actual measurement they are replaced by finite frequency and angular windows. 
For notational simplicity, we write \(h_\lambda\) for \([h_\lambda]_\Delta\) throughout this section, leaving the corresponding resolution windows implicit when comparing with conventional exact-mode spectra. 

Several quantities have also been used in the literature to characterize SGWB anisotropies, including the polarization matrix, Stokes parameters, the direction-dependent strain power spectrum, energy density, and the graviton distribution function.
Here, we clarify their relation to the locally averaged intensity \eqref{eq:intensity SM}. 
Because SGWB anisotropies $\delta_\lambda^{\rm GW}$ are fractional deviations from direction-averaged quantities, direction-independent factors cancel.
Accordingly, in the proportionality relations below, we display the frequency $f$ (or $q$) and sky-direction $\hn$ dependences explicitly, while omitting numerical normalizations and homogeneous background factors.
Frequency-only factors are retained to clarify the correspondence between different spectral conventions. 

First, from the amplitude $h_\lambda$, we can construct the polarization matrix
\begin{align}
J_{\lambda\lambda'}
=
h_\lambda^\dagger(f,\hn) h_{\lambda'}(f, \hn)\,.
\end{align}
In the circular polarization basis, it is related to the Stokes parameters by
\begin{align}
J_{\lambda\lambda'}
=
\frac{1}{2}
\begin{pmatrix}
I+V & Q+iU\\
Q-iU & I-V
\end{pmatrix}.
\end{align}
This matrix contains the same polarization information as Eq.~(A.15) of Ref.~\cite{Romano:2016dpx}.
The signs of $U$ in the off-diagonal components are interchanged because we adopt the ordering $h_\lambda^\dagger h_{\lambda'}$, consistently with the convention in the Letter. 
The helicity-diagonal treatment in the Letter retains the Stokes parameters $I$ and $V$, whereas $Q$ and $U$, which characterize linear polarization, reside in the off-diagonal components. 
The trace component ${\mathcal P}_+ + {\mathcal P}_-$ is identified with the Stokes-$I$ sector, 
whereas the difference ${\mathcal P}_+ - {\mathcal P}_-$ gives the Stokes-$V$ sector. 
We will show that $Q$ and $U$ are generally suppressed in Section \ref{sec:rotational constraints}. 

A stochastic background is theoretically specified by a probability distribution rather than by a particular realization of the GW amplitudes.
Consequently, an individual realization of $J_{\lambda\lambda'}$ cannot be predicted; only its statistical properties can be.
As noted in Ref.~\cite{Romano:2016dpx}, the theoretical polarization matrix of an SGWB is therefore described by a quadratic expectation value $\langle h_\lambda^\dagger h_{\lambda'} \rangle$. Although an individual realization of $J_{\lambda\lambda'}$ is observable, an appropriate average is indispensable for comparing it with a theoretical prediction. 
The quadratic product constructed from the GW fields in our Universe is an observable estimator of this expectation value. 
For an anisotropy measurement, this average must be constructed locally from unresolved short modes within each angular and frequency resolution element.
Perturbations coherent across the element must remain fixed.
Accordingly, the relevant polarization matrix is
    \begin{align}
        \langle
        h_\lambda^\dagger
        h_{\lambda'}
        \rangle_{S,\phi_L} \,,
    \end{align}
where the long-mode realization $\phi_L$ is held fixed.
Its direction-dependent bias constitutes the observable SGWB anisotropy.
The remaining average over $\phi_L$ is performed only when computing correlations between these anisotropies. 

Next, we relate the intensity \eqref{eq:intensity SM} to the standard power spectrum in Fourier space: 
    \begin{align}\label{eq:power spectrum}
         \langle h_{\lambda}^\dagger(\bq) h_{\lambda'}(\bq') \rangle = 
         P_h(q) \delta_{\lambda\lambda'} (2\pi)^3\delta^{(3)}(\bq-\bq') \,.
    \end{align}
The momentum-conserving delta function follows from statistical homogeneity, 
while statistical isotropy implies that $P_h$ depends only on the magnitude $q$. 
These symmetries need not hold for the conditional average at a fixed realization of $\phi_L$.
At linear order, a long mode with wave vector $\bk_L$ correlates short modes whose momenta differ by $\bk_L$; the corresponding contribution has support on
    \begin{align}\label{eq:delta function}
        (2\pi)^3 \delta^{(3)}(\bq-\bq'+\bk_L) \,,
    \end{align}
rather than on $\delta^{(3)}(\bq-\bq')$. The displacement of the momentum-conserving delta function represents the breaking of statistical homogeneity by the fixed long-mode realization. 
Likewise, the corresponding coefficient can depend on the directions of $\bq, \bq'$ relative to $\bk_L$ and on the helicities, reflecting the breaking of statistical isotropy.
Both symmetries are restored after the ensemble average over the long modes.

For the long modes relevant here, however, the displacement $|\bq_2-\bq_1|=k_L$ is smaller than the momentum resolution of the observational window. 
The two short modes therefore lie within the same resolution element, and off-diagonal correlations do not appear in the filtered observable. 
Strictly speaking, for the filtered observable, the delta function \eqref{eq:delta function} is convolved with the observational windows and becomes a finite-width kernel. 
To keep the transformation between the Fourier and frequency--direction representations transparent, we continue to write ideal delta functions with this finite-resolution smearing understood. 
Thus, the biased power spectrum is written as
    \begin{align}\label{eq:anisotropic power spectrum}
        \langle h_{\lambda}^\dagger(\bq) h_{\lambda'}(\bq') \rangle_{S,\phi_L}
         =  
         P_{h, \lambda}(\bq)\delta_{\lambda \lambda'}(2\pi)^3\delta^{(3)}(\bq - \bq') \,,
    \end{align}
where we retain only the helicity-diagonal components. The direction and helicity dependences of
$P_{h,\lambda}(\bq)$ represent the breaking of statistical isotropy by the fixed long-mode realization. 

To connect Eq.~\eqref{eq:anisotropic power spectrum} to the frequency--direction representation, we decompose each Fourier mode into its positive- and negative-frequency components as in Eq.~\eqref{eq:frequency_branches}. 
The two-point function then contains the same-branch terms
\begin{align}
\langle
h_{\lambda}^{(p)\dagger}(\bq)
h_{\lambda'}^{(p)}(\bq')
\rangle \,,
\qquad
\langle
h_{\lambda}^{(n)\dagger}(\bq)
h_{\lambda'}^{(n)}(\bq')
\rangle \,,
\end{align}
and the cross-branch terms
\begin{align}
\langle
h_{\lambda}^{(p)\dagger}(\bq)
h_{\lambda'}^{(n)}(\bq')
\rangle ,
\qquad
\langle
h_{\lambda}^{(n)\dagger}(\bq)
h_{\lambda'}^{(p)}(\bq')
\rangle \,.
\end{align}
Statistical homogeneity requires each of these correlators to be proportional to
$\delta^{(3)}(\bq-\bq')$.
Denoting the coefficients of their helicity-diagonal components by
$P_{h,\lambda}^{(ab)}(\bq)$ with $a,b=p,n$, the power spectrum of the full spatial Fourier mode is decomposed as
\begin{align}\label{eq:Fourier-power-decomposition}
P_{h,\lambda}(\bq)
=
P_{h,\lambda}^{(pp)}(\bq)
+
P_{h,\lambda}^{(nn)}(\bq)
+
P_{h,\lambda}^{(pn)}(\bq)
+
P_{h,\lambda}^{(np)}(\bq) \,.
\end{align}
The first two terms are the same-branch powers, whereas the last two are the cross-branch powers.
These four sectors must be treated separately when translating the Fourier-space power spectrum into the frequency--direction representation.

We first consider the cross-branch terms.
The positive-frequency component with wave vector $\bq$ corresponds to
$f=q/(2\pi)$ and $\hn=-\hq$, whereas the negative-frequency component corresponds to
$f=-q/(2\pi)$ and $\hn=\hq$, with the helicity reversal specified in Eq.~\eqref{eq:Fourier_frequency_dictionary}.
Consequently, for the cross-branch terms, the Fourier-space delta function becomes
\begin{align}\label{eq:cross-branch-delta}
(2\pi)^3\delta^{(3)}(\bq-\bq')
\longrightarrow
\frac{1}{|f|^2}\,
\delta(f+f')\,
\delta^{(2)}(\hn,-\hn') \,.
\end{align}
Thus, $P_{h,\lambda}^{(pn)}$ and $P_{h,\lambda}^{(np)}$ describe correlations between opposite signed frequencies and antipodal sky directions, together with the corresponding helicity reversal.
These are precisely the antipodal correlations discussed in Ref.~\cite{Wu:2022kld}.
As explained in Sec.~\ref{sec:ffn relations}, they are phase sensitive and do not survive the unavoidable frequency and angular coarse graining.
We therefore omit the cross-branch sectors in the remainder of this work.

We next consider the same-branch terms.
For the $pp$ sector, both frequencies are positive, while for the $nn$ sector, both are negative.
In either case, the Fourier-space delta function becomes
\begin{align}\label{eq:same-branch-delta}
(2\pi)^3\delta^{(3)}(\bq-\bq')
\longrightarrow
\frac{1}{|f|^2}\,
\delta(f-f')\,
\delta^{(2)}(\hn,\hn') \,.
\end{align}
Their relation to the Fourier-space branch powers is most transparent in terms of a double-sided spectrum \eqref{eq:aPh SM}, which is defined both for $f>0$ and $f<0$. 
Using Eq.~\eqref{eq:Fourier_frequency_dictionary} and Eq.~\eqref{eq:same-branch-delta}, we obtain
\begin{align}\label{eq:double-sided-dictionary}
\mathcal P_{\lambda}^{\mathrm{(D)}}(f,\hn)
=
\begin{cases}
f^2
P_{h,\lambda}^{(pp)}(-2\pi |f|\hn)
& (f>0) \,,
\\ 
f^2
P_{h,-\lambda}^{(nn)}(2\pi |f|\hn)
& (f<0) \,.
\end{cases}
\end{align}
The reality condition relates the two branches as
\begin{align}
P_{h,\lambda}^{(nn)}(\bq)
=
P_{h,\lambda}^{(pp)}(-\bq),
\end{align}
which is equivalently expressed in the frequency--direction representation as
\begin{align}\label{eq:reality condition}
\mathcal P_{\lambda}^{\mathrm{(D)}}(f,\hn)
=
\mathcal P_{-\lambda}^{\mathrm{(D)}}(-f,\hn).
\end{align}
The positive- and negative-frequency parts of the double-sided spectrum therefore represent the same physical degrees of freedom.

The one-sided spectral density is defined by Eq.~\eqref{eq:one-sided-definition}. 
Hence, the direction-dependent one-sided power spectrum in
Eq.~\eqref{eq:aPh SM} is related to the Fourier-space branch powers as
\begin{align}\label{eq:one-sided-dictionary}
\mathcal P_{\lambda}(f,\hn)
&=
f^2
\left[
P_{h,\lambda}^{(pp)}(-2\pi f\hn)
+
P_{h,\lambda}^{(nn)}(2\pi f\hn)
\right]
\nonumber\\
&=
2f^2P_{h,\lambda}^{(pp)}(-2\pi f\hn)
=
2f^2P_{h,\lambda}^{(nn)}(2\pi f\hn).
\end{align}
Setting $\bq_\ast \equiv -2\pi f\hn$ for given $f>0$ and $\hn$, 
\begin{align}
\frac{\mathcal P_{\lambda}(f,\hn)}{f^2}
=
P_{h,\lambda}^{(pp)}(\bq_\ast)
+
P_{h,\lambda}^{(nn)}(-\bq_\ast)
=
2P_{h,\lambda}^{(pp)}(\bq_\ast) \,,
\end{align}
while
\begin{align}
\left. P_{h,\lambda}(\bq_\ast)
\right|_{\rm no\ cross\ terms}
=
P_{h,\lambda}^{(pp)}(\bq_\ast)
+
P_{h,\lambda}^{(nn)}(\bq_\ast)
=
P_{h,\lambda}^{(pp)}(\bq_\ast)
+
P_{h,\lambda}^{(pp)}(-\bq_\ast) \,.
\end{align}
Thus, the two quantities carry different information unless 
\begin{align}\label{eq:standing wave}
P_{h,\lambda}^{(pp)}(\bq)
=
P_{h,\lambda}^{(pp)}(-\bq) \,.
\end{align}
This condition is satisfied by a standing-wave configuration, for
which the two counter-propagating components carry equal power.
In particular, it is realized when the subhorizon solution is matched to the constant superhorizon mode, as in the primordial cosmological backgrounds~\cite{Wu:2022kld}.

Many studies of the cosmological SGWB characterize anisotropies in
terms of a GW energy-density spectrum rather than the strain power spectrum introduced above. 
For freely propagating GWs on subhorizon scales, the local GW energy density is given by
\begin{align}
\rho_{\rm GW}(t,\bx)
\propto
\overline{
\partial_t h_{ij}(t,\bx)
\partial_t h_{ij}(t,\bx)
} \,,
\label{eq:local GW energy density}
\end{align}
where the overbar denotes a temporal average over the rapid GW oscillations. 
The local quadratic quantity in Eq.~\eqref{eq:local GW energy density} does not yet define a directional energy density. 
Expanding Eq.~\eqref{eq:local GW energy density} in the frequency--direction basis gives
\begin{align}
\overline{
\partial_t h_{ij}(t,\bx)\partial_t h_{ij}(t,\bx)
}
&\propto
\sum_{\lambda,\lambda'}
\int\rd f\,\rd f'
\int\rd^2\hn\,\rd^2\hn'
\,ff'
e^\lambda_{ij}(-\hn) e^{\lambda'}_{ij}(-\hn') \nonumber\\
&\qquad\qquad \times
\overline{
e^{-2\pi i(f+f')t}
}
e^{-2\pi i[f\hn+f'\hn']\cdot\bx}
h_\lambda(f,\hn)
h_{\lambda'}(f',\hn') \,.
\end{align}
For an averaging time much longer than the periods of the GWs, 
the temporal average restricts the integral to $f'\simeq-f$. 
Using the reality condition, 
the surviving quadratic combination can be written as
    \begin{align}
        h_\lambda(f,\hn)h_{\lambda'}(-f,\hn')
        =
        h_\lambda(f,\hn) h_{-\lambda'}^\dagger(f,\hn') \,.
    \end{align}
The time average alone, however, does not require $\hn'=\hn$: 
interference between different directions at the same frequency may remain. 
Reducing the two angular integrals to a single one therefore requires an additional spatial coarse graining over the position $\bx$, 
    \begin{align}\label{eq:spatial average}
        \overline{e^{-2\pi i f (\hn-\hn')\cdot\bx}} \propto \delta^{(2)}(\hn, \hn') \,,
    \end{align}
or a statistical average, 
    \begin{align}\label{eq:GW average}
    \langle h_\lambda^\dagger(f,\hn) h_{\lambda'}(f',\hn') \rangle_{S,\phi_L}
    \propto
    \delta^{(2)}(\hn,\hn') \,.
    \end{align}
Here, the overbar denotes a spatial coarse graining over $\bx$, in contrast to the brackets used for statistical averages over the GW field. 
The former coarse graining is implicitly supplied by the finite spatial resolution with which the local energy density is assigned to a position along a line of sight. 
The latter conditional statistical average may be approximated observationally by combining suitable quadratic estimators within each angular/frequency resolution element, over repeated time segments, and independent detector baselines. 

After either of these procedures, the surviving phase-insensitive contribution can be resolved into a direction-dependent energy-density spectrum, 
    \begin{align} \label{eq:egw}
    {\cal E}_{\rm GW}(f,\hn)
    \equiv
    \frac{\rd^3\rho_{\rm GW}}
    {\rd\ln f\,\rd^2\hn} \,.
    \end{align}
Here and below, the spectral energy density is defined for $f>0$, following the standard one-sided convention; 
the negative-frequency branch appearing in the preceding signed-frequency expansion is included through the reality relation \eqref{eq:reality relation}. 
The energy-density spectrum \eqref{eq:egw} defined with the statistical average \eqref{eq:GW average} is directly related to the direction-dependent strain power spectrum by
\begin{align}
{\cal E}_{\rm GW}(f,\hn)
\propto
f^3{\cal P}(f,\hn)
=
f^3\sum_\lambda{\cal P}_\lambda(f,\hn).
\label{eq:directional GW energy density}
\end{align}
The spatial coarse graining \eqref{eq:spatial average} yields a realization-dependent estimator and may require the additional averaging procedure to obtain the relation \eqref{eq:directional GW energy density}. 
Integrating over the sky gives the usual present-day energy-density spectrum,
\begin{align}
\Omega_{\rm GW}(f)
\equiv
\frac{1}{\rho_{\rm c}}
\frac{\rd\rho_{\rm GW}}{\rd\ln f}
\propto
f^3S_h(f).
\label{eq:GW energy density}
\end{align}
At fixed frequency, all the conversion factors in
Eq.~\eqref{eq:directional GW energy density} are independent of $\hn$.
The fractional energy-density anisotropy is therefore the
polarization-averaged strain-power anisotropy,
\begin{align}\label{eq:pol av delta}
\delta^{\rm GW}(f,\hn)
\equiv
\frac{1}{2}\sum_\lambda
\delta_\lambda^{\rm GW}(f,\hn) \,.
\end{align}

Finally, in the geometric-optics regime, the direction-resolved energy density can be equivalently represented by a graviton distribution function. 
Denoting the latter by $f_{\rm GW}(\tau,\bx,\bq)$, 
the local energy density is written as
\begin{align}
\rho_{\rm GW}(\tau,\bx)
\propto
\int\rd^3\bq\,
q\,f_{\rm GW}(\tau,\bx,\bq),
\end{align}
where $\bq=q\hq$ is the comoving graviton momentum and $\hq$ denotes
its propagation direction. Factors independent of $q$ and $\hq$,
including the appropriate powers of the scale factor, are left
implicit according to the convention adopted above. 
Using $\rd^3\bq=q^3\rd\ln q\,\rd^2\hq$, 
we obtain
\begin{align}\label{eq:fgw}
f_{\rm GW}(\tau,\bx,q,\hq)
\propto
q^{-4}
\frac{\rd^3\rho_{\rm GW}}
{\rd\ln q\,\rd^2\hq} \,.
\end{align}
The conversion factor is independent of direction. 
The distribution function \eqref{eq:fgw} and the direction-dependent energy-density spectrum \eqref{eq:egw} therefore
contain the same fractional angular information at each fixed $q$.
To translate this phase-space convention into the sky-direction
convention used in the Letter, one should identify
\begin{align}
q=2\pi f,
\qquad
\hq=-\hn \qquad (f>0) \,.
\end{align}
The distribution-function description applies to the phase-insensitive,
direction-diagonal sector obtained after the coarse graining discussed above. 
It does not retain the phase-sensitive cross-branch information, such as the antipodal correlations, which cannot be represented by a classical graviton occupation number.

Closely related constructions appear in the cosmological-GW literature, although the directional label and the associated coarse graining are introduced in different ways.  
Ref.~\cite{Bartolo:2019zvb} first constructs a position-dependent energy density $\rho_{\rm GW}(\tau,\bx)$, or equivalently $\Omega_{\rm GW}(\tau,\bx,f)$, and obtains a sky map by evaluating this local quantity at the source position associated with each line of sight: $\rho(t, \bx)|_{\bx=\hn(\tau_0-\tau)}$. 
For a map with finite angular and radial resolution, the value assigned to each line of sight is implicitly coarse grained over a finite region around the corresponding source position.  
This spatial coarse graining plays the role of the $\bx$ average discussed around Eq.~\eqref{eq:spatial average}. 
Refs.~\cite{Dimastrogiovanni:2021mfs,Li:2023xtl,Li:2023qua}, on the other hand, formulate the energy-density spectrum directly in terms of the short-scale Fourier power spectrum or a graviton phase-space distribution.  
In particular, Refs.~\cite{Li:2023xtl,Li:2023qua} introduce a position-dependent spectrum whose integral over the magnitude and direction of the short GW momentum gives the local energy density.  
The direction of this momentum is subsequently interpreted as the graviton propagation direction in the Boltzmann description; the unit vector associated with the short GW momentum in these references denotes the propagation direction and is therefore opposite to the sky direction $\hn$ adopted here.
In these formulations, the direction-diagonal short-mode power spectrum is obtained through a statistical average, explicitly or implicitly, corresponding to the second route in Eq.~\eqref{eq:GW average}.  
A temporal average may additionally remove the rapid GW oscillations, but it does not eliminate interference between different directions, as discussed above.

Viewed from the present framework, two qualifications are important.  
First, these formulations do not explicitly distinguish the positive- and negative-frequency branches.  
As demonstrated in Eqs.~\eqref{eq:Fourier-power-decomposition} and \eqref{eq:one-sided-dictionary}, a decomposition by the spatial Fourier direction is not generally identical to a decomposition by the observed arrival direction.  
The distinction becomes immaterial under the equal-power condition \eqref{eq:standing wave}, which is realized for the primordial standing-wave configurations relevant to the references above.  
Second, the statistical average implicit in a Fourier power spectrum or distribution function must be distinguished from the full ensemble average.  
The quantity relevant to the CFF is the conditional average $\langle\cdot\rangle_{S,\phi_L}$ over the short sector at a fixed long-mode realization; averaging also over $\phi_L$ would restore statistical homogeneity and isotropy and erase the anisotropy of interest.  
Spatial coarse graining suppresses direction-off-diagonal interference but leaves a realization-dependent quadratic quantity. It estimates the conditional power spectrum only when the modes combined by the observational window can be treated as statistically equivalent samples of the short-mode distribution at fixed $\phi_L$. 
For an astrophysical background, 
such sampling may be provided by statistically independent time segments or incoherent sources. 
For a primordial background, 
self-averaging may instead be achieved by combining short Fourier modes within frequency and angular bins. 
Combining modes with different frequencies or directions requires theoretical templates for their frequency and angular dependence within the bins. 
The CFF provides the angular-response template, while its process-dependent frequency dependence must be specified or inferred separately. 

In summary, the direction-dependent strain power, the directional
energy-density spectrum, and the graviton distribution function provide equivalent representations of the same locally averaged, phase-insensitive sector of the correlations \eqref{eq:intensity SM}. 
This equivalence should be understood with the direction conventions, standing-wave condition, and averaging prescriptions specified above. 
The CFF framework makes the underlying conditional average explicit and encodes the response of these short-mode statistics to the fixed long modes that generate their anisotropies.


\section{Long--Short Mode Splitting and the CFF--Bispectrum Relation}
\label{sec:SM_splitting}
\label{sec:derivations}

In this section, we make the long--short mode separation more precise and provide a detailed derivation of the relation between the Cosmological Form Factor (CFF), the SGWB anisotropy, and the full-ensemble three-point function.
The distinction between long and short modes depends on the resolution and averaging procedure of the observation rather than being an intrinsic property of the perturbations themselves.

We begin by introducing the conditional probability framework that underlies the response interpretation.

\subsection{Conditional probability and the response to long modes}

The fundamental object in our formalism is the conditional probability distribution $\mathbb{P}[h_{\Delta} | \phi_L]$, which describes the statistics of the short-wavelength gravitational wave field $h_{\Delta}$ (smoothed over small scales) given a fixed realization of the long-wavelength perturbation $\phi_L$. 
This conditional distribution encodes how the long modes bias the local observables. 
Closely related conditional-response methods have been used in the study of large-scale structure, known as the ``clustering fossil'' effect~\cite{wang2024reconstructing}.

The joint probability distribution of the short and long modes can be written as
\begin{align}\label{eq:joint P}
\mathbb{P}[h_\Delta,\phi_L]
=
\mathbb{P}[h_\Delta|\phi_L]\mathbb{P}[\phi_L] \,.
\end{align}
Marginalizing over the long modes gives the unconditional probability distribution of the short modes,
\begin{align}
\mathbb{P}[h_{\Delta}] = \int \mathcal{D}\phi_L \, \mathbb{P}[h_{\Delta} | \phi_L] \, \mathbb{P}[\phi_L] \,.
\end{align}
For any observable $\mathcal{O}[h_\Delta]$ that depends only on the short modes, the law of total expectation gives
\begin{align}\label{eq:total expectation}
\langle \mathcal{O}[h_{\Delta}] \rangle = \left\langle \langle \mathcal{O}[h_{\Delta}] \rangle_{S,\phi_L} \right\rangle_L \,,
\end{align}
where $\langle \cdot \rangle_{S,\phi_L}$ denotes the expectation value with respect to $\mathbb{P}[h_{\Delta} | \phi_L]$ (i.e., averaging over short modes with fixed long modes), and $\langle \cdot \rangle_L$ denotes averaging over $\mathbb{P}[\phi_L]$.

The key step in formulating the response problem is to regard the conditional expectation value introduced in Eq.~\eqref{eq:total expectation} as a functional of the long-mode configuration $\phi_L$. 
Treating the amplitude of $\phi_L$ perturbatively, we expand the conditional expectation value around $\phi_L=0$. 
At linear order, we have
\begin{align}\label{eq:response x}
\langle \mathcal{O}[h_{\Delta}] \rangle_{S,\phi_L} 
= \langle \mathcal{O}[h_{\Delta}] \rangle_{S,\phi_L=0} + \int d^3\bx \, \frac{\delta \langle \mathcal{O}[h_{\Delta}] \rangle_{S,\phi_L}}{\delta \phi_L(\bx)} \bigg|_{\phi_L=0} \phi_L(\bx) + {\cal O}(\phi_L^2) \,,
\end{align}
in real space, or
\begin{align}\label{eq:response k}
\langle \mathcal{O}[h_{\Delta}] \rangle_{S,\phi_L} 
= \langle \mathcal{O}[h_{\Delta}] \rangle_{S,\phi_L=0} + \int \rd^3\bk_L \, \frac{\delta \langle \mathcal{O}[h_{\Delta}] \rangle_{S,\phi_L}}{\delta \phi(\bk_L)} \bigg|_{\phi_L=0} \phi(\bk_L) + {\cal O}(\phi_L^2)  \,,
\end{align}
in Fourier space. 
Here, $\langle \mathcal{O}[h_{\Delta}] \rangle_{S,\phi_L=0}$ is the unmodulated expectation value. 
For the local intensity $\mathcal{O}[h_{\Delta}] = |[h_{\lambda}]_{\Delta}(f,\hn)|^2$, it is identified with the usual isotropic short-mode power spectrum, 
and the expansion \eqref{eq:response k} takes the form
\begin{align}\label{eq:CFF SM}
\left\langle |[h_{\lambda}]_{\Delta}(f,\hn)|^2 \right\rangle_{S,\phi_L} 
= \left\langle |[h_{\lambda}]_{\Delta}(f,\hn)|^2 \right\rangle_{S,\phi_L=0} \left[ 1 + \int \frac{\rd^3\bk_L}{(2\pi)^3} F_{\lambda}(f,\hn,\bk_L) \phi(\bk_L) \right] \,,
\end{align}
where
\begin{align}
    F_{\lambda}(f,\hn,\bk_L)
    \equiv 
    \frac{(2\pi)^3}{\left\langle |[h_{\lambda}]_{\Delta}(f,\hn)|^2 \right\rangle_{S,\phi_L=0}} \frac{\delta \langle \mathcal{O}[h_{\Delta}] \rangle_{S,\phi_L}}{\delta \phi(\bk_L)} \bigg|_{\phi_L=0} \,.
\end{align}
The coefficient $F_{\lambda}(f,\hn,\bk_L)$ is precisely the Cosmological Form Factor (CFF), which encodes the linear response of the local short-mode intensity to a long-wavelength perturbation.

\subsection{Power spectrum of the smoothed field}

Substituting the definition of the filtered gravitational-wave amplitude, the smoothed power is related to the unsmoothed one as
\begin{align}
\left\langle |[h_{\lambda}]_{\Delta}(f,\hn)|^2 \right\rangle
= \int\rd f_1\rd f_2
\int\rd^2\hn_1\rd^2\hn_2\,
W_{\Delta}^*(f,\hn;f_1,\hn_1)
W_{\Delta}(f,\hn;f_2,\hn_2)
\left\langle
h_{\lambda}^\dagger(f_1,\hn_1)
h_{\lambda}(f_2,\hn_2)
\right\rangle \,.
\end{align}
For a statistically homogeneous and isotropic unpolarized SGWB, 
\begin{align}\label{eq:smoothed ps}
\left\langle |[h_{\lambda}]_{\Delta}(f,\hn)|^2 \right\rangle_{S,\phi_L=0}
\simeq \frac{S_h(f)}{16\pi}
\int\rd f'\int\rd^2\hn'\,
\left|W_{\Delta}(f,\hn;f',\hn')\right|^2
\equiv \frac{S_h(f)N_{\Delta}}{16\pi} \,,
\end{align}
where we have assumed that $S_h(f')$ varies negligibly across the frequency window. 
Using Eq.~\eqref{eq:smoothed ps} in Eq.~\eqref{eq:CFF SM}, we obtain
\begin{align}\label{eq:CFF SM2}
\left\langle |[h_{\lambda}]_{\Delta}(f,\hn)|^2 \right\rangle_{S,\phi_L}
= 
\frac{S_h(f)N_\Delta}{16\pi}\left[ 1 + \int \frac{\rd^3\bk_L}{(2\pi)^3} F_{\lambda}(f,\hn,\bk_L) \phi(\bk_L) \right] \,.
\end{align}
This is the CFF relation given in the Letter.

\subsection{Relation to the three-point function}

To express the CFF in terms of a three-point correlation function, we consider the full average over both short and long modes. 
Using the law of total expectation ({\it i.e.}, averaging first over short modes conditioned on the long mode, then over long modes) [See Eq.~\eqref{eq:joint P}], 
\begin{align}
\left\langle |[h_{\lambda}]_{\Delta}(f,\hn)|^2 \phi^\dagger(\bk_L) \right\rangle 
= \left\langle
\left\langle |[h_{\lambda}]_{\Delta}(f,\hn)|^2 \right\rangle_{S,\phi_L}
\phi^\dagger(\bk_L)
\right\rangle_L \,.
\end{align}
Substituting the linear response expression \eqref{eq:CFF SM2}, we find
\begin{align}
\left\langle |[h_{\lambda}]_{\Delta}(f,\hn)|^2 \phi^\dagger(\bk_L) \right\rangle
= \frac{S_h(f)N_{\Delta}}{16\pi}
\left\langle \phi^\dagger(\bk_L) \right\rangle_L 
+ \frac{S_h(f)N_{\Delta}}{16\pi}
\int \frac{\rd^3\bk_L'}{(2\pi)^3}
F_{\lambda}(f,\hn,\bk_L')
\left\langle \phi(\bk_L')\phi^\dagger(\bk_L)\right\rangle_L \,.
\end{align}
The first term vanishes because $\langle \phi^\dagger(\bk_L)\rangle_L=0$. 
The second term involves the power spectrum of $\phi_L$, which is defined by
\begin{align}
\left\langle \phi(\bk_L')\phi^\dagger(\bk_L)\right\rangle_L
=(2\pi)^3\delta^{(3)}(\bk_L'-\bk_L)P_\phi(k_L) \,.
\end{align}
Performing the integration over $\bk_L'$, we obtain
\begin{align}
\left\langle |[h_{\lambda}]_{\Delta}(f,\hn)|^2 \phi^\dagger(\bk_L) \right\rangle 
=\frac{S_h(f)N_{\Delta}}{16\pi}
F_{\lambda}(f,\hn,\bk_L)P_\phi(k_L) \,.
\end{align}
Thus,
\begin{align}\label{eq:CFF B SM}
F_{\lambda}(f,\hn,\bk_L)
=\frac{16\pi}{S_h(f)P_\phi(k_L)N_\Delta}
\left\langle |[h_{\lambda}]_{\Delta}(f,\hn)|^2
\phi^\dagger(\bk_L)\right\rangle \,.
\end{align}
This is the full-ensemble three-point-function relation given in the Letter.

\subsection{Relation to the Fourier-space bispectrum}

We next relate the three-point function in Eq.~\eqref{eq:CFF B SM} to the Fourier-space bispectrum. 
The spatial Fourier representation makes statistical homogeneity manifest and identifies the range of long modes that can contribute to the CFF. 
It is also useful for analyzing the polarization structure in Section \ref{sec:rotational constraints}. 

Using the correspondence in Eq.~\eqref{eq:Fourier_frequency_dictionary}, the filtered amplitude \eqref{eq:filtered amplitude} for $f>0$ can be mapped to its spatial-Fourier counterpart as 
\begin{align}
[h_\lambda]_\Delta(f,\hn)
=
[h^{(p)}_\lambda]_\Delta(-2\pi f\hn) \qquad (f>0) \,.
\label{eq:Fourier_frequency_dictionary filtered}
\end{align}
Here, we have defined 
\begin{align}\label{eq:smoothed GW field I}
[h_{\lambda}^{(p)}]_{\Delta}(\bq) 
\equiv 
\int\frac{\rd^3\bp}{(2\pi)^3}\,
\widetilde W_{\Delta}(\bq,\bp)
h^{(p)}_{\lambda}(\bp) \,,
\end{align}
where normalization factors have been absorbed into the induced Fourier-space window $\widetilde W_\Delta$. 
Using the mapping \eqref{eq:Fourier_frequency_dictionary filtered}, 
the three-point function in Eq.~\eqref{eq:CFF B SM} can be rewritten as
    \begin{align}
        \left\langle |[h_{\lambda}]_{\Delta}(f,\hn)|^2 \phi^\dagger(\bk_L)\right\rangle 
        =
         \left\langle |[h_{\lambda}^{(p)}]_{\Delta}(\bq)|^2 \phi^\dagger(\bk_L)\right\rangle \qquad (\bq = -2\pi f \hn) \,.
    \end{align}
Substituting Eq.~\eqref{eq:smoothed GW field I} gives 
\begin{align}
\left\langle |[h_{\lambda}^{(p)}]_{\Delta}(\bq)|^2
\phi^\dagger(\bk_L)\right\rangle
=
\int\frac{\rd^3\bq_1\rd^3\bq_2}{(2\pi)^6}\,
\widetilde W_{\Delta}^*(\bq,\bq_1)
\widetilde W_{\Delta}(\bq,\bq_2) 
\left\langle
h^{(p)\dagger}_{\lambda}(\bq_1)
h^{(p)}_{\lambda}(\bq_2)
\phi^\dagger(\bk_L)
\right\rangle \,.
\end{align}
Thus, the CFF probes the nearly collinear configuration $\bq_1\simeq\bq_2\simeq\bq$ of the Fourier-space bispectrum
\begin{align}\label{eq:3pt I}
\left\langle
h^{(p)\dagger}_{\lambda_1}(\bq_1)
h^{(p)}_{\lambda_2}(\bq_2)
\phi^\dagger(\bk_L)
\right\rangle \,.
\end{align}
Statistical homogeneity requires
\begin{align}\label{eq:bispectrum_SM}
\left\langle
h^{(p)\dagger}_{\lambda_1}(\bq_1)
h^{(p)}_{\lambda_2}(\bq_2)
\phi^\dagger(\bk_L)
\right\rangle 
=
B_{\lambda_1\lambda_2}(\bq_1,\bq_2,\bk_L)
(2\pi)^3\delta^{(3)}(-\bq_1+\bq_2-\bk_L) \,.
\end{align}
A convenient parametrization of the momentum-conservation condition is
\begin{align}
\bq_1=\bq+\delta\bq_1 \, ,\qquad
\bq_2=\bq+\delta\bq_2 \, ,\qquad
\bk_L=\delta\bq_2-\delta\bq_1 \,.
\end{align}
Consequently, the CFF receives contributions only from long modes whose wave vectors lie within the momentum resolution of the short-mode window.
For a window with frequency width $\Delta f$ and angular width $\Delta\theta_{\rm sm}$, the corresponding radial and transverse resolutions are
\begin{align}
|\Delta q_\parallel|\sim2\pi\Delta f \, ,
\qquad
|\Delta\bq_\perp|\sim 2\pi f\Delta\theta_{\rm sm} \,.
\end{align}
When the angular resolution determines the characteristic momentum width and $\Delta\theta_{\rm sm}\sim\ell_{\max}^{-1}$, this gives
\begin{align}\label{eq:long mode kinematic}
k_L\lesssim |\Delta\bq|
\sim 2\pi f\Delta\theta_{\rm sm}
\sim 2\pi f\ell_{\max}^{-1} \,.
\end{align}
This condition is a kinematic consequence of the smoothing procedure and is independent of locality.

Upon imposing locality, the support of the real-space response kernel in Eq.~\eqref{eq:response x} is restricted to the angular patch selected by the smoothing window. 
The resulting anisotropy can be written schematically as
\begin{align}
\delta^\text{GW}_\lambda(f,\hn)
=\int\frac{\rd^3\bk_L}{(2\pi)^3}
\left[
\int_{\rm patch}\rd^3\bx\,
F_\lambda(f,\hn,\bx)e^{i\bk_L\cdot\bx}
\right]\phi(\bk_L) \,.
\end{align}
The patch is centered on the observed sky direction $\hn$ with angular size $\Delta\theta_{\rm sm}\sim\ell_{\max}^{-1}$. 
Up to radial measure factors, the integration can be written as
\begin{align}
\int_{\rm patch}\rd^3\bx
=\int\rd\tau_s
\int_{|\hn'-\hn|<\Delta\theta_{\rm sm}}\rd^2\hn' \, ,
\qquad
\bx=\hn'(\tau_0-\tau_s) \,.
\end{align}
Within the local patch, the phase factor $e^{i\bk_L\cdot\bx}$ should remain approximately constant; otherwise, the oscillatory integral averages out. This gives
\begin{align}
k_L\Delta\theta_{\rm sm}(\tau_0-\tau_s)\lesssim1 \,.
\end{align}
For sources at cosmological distance, $\tau_s\ll\tau_0$, and hence
\begin{align}
k_L\tau_0\lesssim\Delta\theta_{\rm sm}^{-1}\sim\ell_{\max} \,.
\end{align}
Combining this locality condition with the momentum-resolution condition, the long modes relevant for the local CFF satisfy
\begin{align}\label{eq:long mode cond}
k_L\lesssim
\min\left(\frac{2\pi f}{\ell_{\max}},\frac{\ell_{\max}}{\tau_0}\right) \,.
\end{align}
Since $2\pi f\tau_0\gg\ell_{\max}^2$ for the frequency range of interest, the locality condition $k_L\tau_0\lesssim\ell_{\max}$ is typically the more restrictive one.

The condition $k_L\tau_0\lesssim\ell_{\max}$ implies the large hierarchy between the long and short modes. 
For space-based gravitational-wave detectors operating around the mHz band, we find
\begin{align}\label{eq:hierarchy}
\frac{k_L}{q}
\lesssim10^{-16}\ell_{\max}
\left(\frac{f}{10^{-3}{\rm Hz}}\right)^{-1} \, ; q = 2\pi f \,,
\end{align}
This enormous hierarchy provides the basis for the conditional-average description adopted in our approach.
The long mode $\phi_L$ and the short-wavelength gravitational-wave field can be regarded as distinct stochastic degrees of freedom, separated by their characteristic wavelengths.
Because the relevant long modes vary negligibly across a local smoothing patch, their realization can be held fixed while the short modes are averaged. 
This directly leads to the conditional expectation value $\langle\cdots\rangle_{S,\phi_L}$. 

\section{Rotational Constraints on the CFF}\label{sec:rotational constraints}

In this section, we will see how symmetries constrain the polarization and angular structures of the CFF. 
It will be shown that the diagonal components $\lambda_1=\lambda_2$ remain at the leading order. 
In the Letter, therefore, we have defined the CFF only for the diagonal components. 

\subsection{Little Group}
The polarization and angular structures of the CFF are determined by the bispectrum $B_{\lambda_1 \lambda_2}(\bq_1, \bq_2, \bk_L)$ in Eq.~\eqref{eq:bispectrum_SM}.  
In the discussion below, 
it is convenient to solve the momentum conservation as,
    \begin{align}
        \bq_1 = \bq_c - \frac{\bk_L}{2} \,, \quad 
        \bq_2 = \bq_c + \frac{\bk_L}{2} \,,
    \end{align}
with $\bq_c \equiv (\bq_1+\bq_2)/2$. 
Since the large hierarchy exists between the long and short modes $k_L/q \ll 1$ [See Eq.~\eqref{eq:hierarchy}], 
the two short-mode directions are approximately collinear, 
    \begin{align}
        \hq_a = \hq_c + {\cal O}(k_L/q) \,,
    \end{align}
and therefore
    \begin{align}\label{eq:poralization basis approx}
        e_{ij}^\pm(\hq_a) = e_{ij}^\pm(\hq_c) + {\cal O}(k_L/q) \,.
    \end{align}

To investigate polarization structures, 
we introduce the residual rotation group $SO(2)_{\bq_c}$, namely the subgroup of spatial rotations leaving $\hq_c$ invariant.
This little group defines the helicity decomposition of the short modes. 
Under a little-group rotation, {\it i.e.,} a rotation by an angle $\theta$ around the $\hq_c$-axis, 
the polarization basis for $\hq_c$ transforms as,
    \begin{align}\label{eq:transformation e}
        e_{ij}^\pm(\hq_c) \to  e^{\pm 2i\theta} e_{ij}^\pm(\hq_c) \,,
    \end{align}
and consequently
\begin{align}
    h^{(p)}_{\lambda_a}(\bq_a) \to e^{2i\lambda_a \theta}h^{(p)}_{\lambda_a}(\bq_a) + {\cal O}(k_L/q) \quad (a=1,2) \,,
\end{align}
with $\lambda_a = \pm$. 

As we mentioned in Section \ref{sec:SM_splitting}, 
the observable filter induces a finite wavevector resolution in the Fourier representation. 
The short-mode wavevectors contributing to the CFF satisfy  
\begin{align}
\bq_a = \bq + \delta\bq_a; \qquad 
|\delta\bq_a| \lesssim q/\ell_{\max} \,,
\end{align}
from which
\begin{align}\label{eq:qc q}
\hq_c = \hq + {\cal O}(\ell_{\max}^{-1}) \,,
\end{align}
where $\bq$ denotes the central wavevector of the smoothing. 
Therefore, the helicity defined with respect to $\hq_c$ coincides with that of the observational polarization basis $e_{ij}^{\lambda}(-\hn)$, since $\hq=-\hn$ for the positive-frequency branch, up to corrections of order $\ell_{\max}^{-1}$. 

\subsection{Rotational Constraints for Scalar Long Modes}

Rotational covariance implies that the bispectrum \eqref{eq:bispectrum_SM} can be decomposed as
    \begin{align}
        B_{\lambda_1\lambda_2}(\bq_1, \bq_2, \bk_L) 
        =
        e_{ij}^{-\lambda_1}(\hq_1) e_{kl}^{\lambda_2}(\hq_2) T^{ijkl} \,,
    \end{align}
where $T^{ijkl}$ is a rotationally covariant rank-four tensor constructed from the available building blocks,
    \begin{align}
        q_1^i\,, \quad q_2^i\,, \quad k_L^i \,, \quad \delta^{ij} \,, \quad \epsilon^{ijk} \,.
    \end{align}
We organize the tensor structure of $T^{ijkl}$ in a systematic expansion in the long-mode wavevector $\bk_L$. 
At leading order ${\cal O}(k_L^0)$, $\bq_a \simeq \bq_c$ and the independent building blocks reduce to
    \begin{align}
        q_c^i\,, \quad \delta^{ij} \,, \quad \epsilon^{ijk} \,. 
    \end{align}
Since each of these building blocks carries only helicity zero under $SO(2)_{\bq_c}$, 
their tensor product $T^{ijkl}$ also contains only helicity-zero components. 
Consequently, its projection onto the polarization basis $e_{ij}^{\lambda_a}(\hq_a) \simeq e_{ij}^{\lambda_a}(\hq_c)$ can only generate the helicity-zero structures corresponding to the intensity (I) and circular polarization (V) sectors,
    \begin{align}
        \delta_{\lambda_1 \lambda_2} \,, \quad \lambda_1 \delta_{\lambda_1 \lambda_2} \,. 
    \end{align}
These two independent helicity-zero quantities correspond to the parity-even and parity-odd structures, respectively. 
The helicity-four combinations associated with the linear polarization $(Q,U)$ are absent at leading order. 
Hence, the bispectrum only has diagonal components $\lambda_1=\lambda_2$. 

To produce the linear polarization $(Q,U)$, 
we need to consider higher-order terms in $\bk_L$. 
At higher orders, the transverse component of the long-mode wavevector,
\begin{align}
    \bk_{L,\perp}
    \equiv
    \bk_L - \hq_c(\hq_c\cdot\bk_L) \,,
\end{align}
provides a helicity-one building block. 
Depending on the underlying physical process, the dimensionless expansion parameter is either $k_L\tau_c$ or $k_L/q$. 
Here, $\tau_c$ denotes the characteristic spacetime scale over which the helicity-carrying response is generated coherently. 
Its physical meaning depends on the underlying mechanism. 
For example, it corresponds to the coherence length along the line of sight for propagation-induced effects. 
Local multipoles of the graviton distribution give helicity-carrying objects, 
which can be transferred to the polarization of the outgoing line-of-sight gravitons 
if gravitons undergo large-angle scatterings analogous to Thomson scattering in the CMB. 
In this case, $\tau_c$ is the characteristic time over which these multipoles are generated by free streaming. 
For modes contributing to the angular multipole $\ell$, $k_L\tau_0\sim\ell$ up to factors of order unity.
Thus, we find $k_L \tau_c \sim \ell (\tau_c/\tau_0) < \ell_{\max} (\tau_c/\tau_0)$. 
Thus, $k_L \tau_c$ is small unless the characteristic scale $\tau_c$ is comparable to the cosmological scale. $k_L/q$ is also extremely small as we have argued. 

By contrast, no such coherence requirement exists for the helicity-zero sector $(I,V)$. Since the helicity-carrying object is a scalar, its response can accumulate coherently over cosmological distances without requiring a fixed orientation in helicity space. This is realized, for example, by the propagation phase $e^{i\hq\cdot\bk_L (\tau_0-\tau_s)}$, whose dependence on $\bk_L$ is not suppressed. 

After the polarization tensors have been projected out,
the remaining coefficient must be a rotational scalar. 
The allowed scalar quantities are
\begin{align}
q_c \equiv |\bq_c|,\qquad k_L \equiv |\bk_L|,\qquad \hq_c\cdot\hk_L \,.
\end{align}
The above constraints have been formulated with respect to the central short-mode wavevector $\bq_c$, which naturally characterizes the squeezed bispectrum. 
As we have argued in Eq.~\eqref{eq:qc q}, 
$\bq_c$ is identical to the observational wavevector $\bq$ up to corrections of order $\ell_{\max}^{-1}$. 
Accordingly, the leading-order polarization structure is unaffected by replacing $(q_c,\hq_c)$ with $(q,\hq)$. 
Expressing the result in terms of the observational Fourier wavevector $\bq=-2\pi f\hn$, 
the polarization and angular structures of the bispectrum are restricted as
    \begin{align}
        B_{\lambda_1\lambda_2}(\bq_1, \bq_2, \bk_L) 
        =
        \delta_{\lambda_1\lambda_2}
        \left[ 
        B_I(q, k_L, \hq\cdot\hk_L) + \lambda_1 B_V(q, k_L,\hq\cdot\hk_L)
        \right] \,,
    \end{align}
where $B_I$ and $B_V$ are associated with the intensity and circular polarization of gravitational waves, respectively. 
Through the relation in Eq.~\eqref{eq:CFF B SM}, the CFF must therefore take the form
\begin{align}
F_{\lambda}(f,\hn,\bk_L) 
&= F_\lambda(f,k_L,\hn\cdot\hk_L) \notag \\
&= F_I(f,k_L,\hn\cdot\hk_L) + \lambda F_V(f,k_L,\hn\cdot\hk_L) \,.
\end{align}
This justifies the angular and polarization structures stated in the Letter.

\section{Recasting existing results in the CFF framework}\label{sec:SM_examples}

SGWB anisotropies have previously been studied using a variety of approaches, including the Boltzmann-equation approach, perturbative diagrammatic calculations, and fossil-type power-spectrum analyses. 
In this section, we illustrate how different anisotropy mechanisms can be interpreted within the common CFF framework for selected examples.

\subsection{Example I: Propagation effects in the Boltzmann-equation approach}

We now illustrate, following the treatment of Bartolo {\it et al.}~\cite{Bartolo:2019yeu}, how SGWB anisotropies induced by propagation effects fit into the CFF formalism. 
The graviton distribution function $f_{\rm GW}(\tau, \bx,\bq)$ obeys the Boltzmann equation
\begin{equation}
\frac{\partial f_{\rm GW}}{\partial \tau}
+\hq\cdot\nabla f_{\rm GW}
+\left[
\frac{\partial\Psi}{\partial\tau}
-\hq\cdot\nabla\Phi
\right]
q\frac{\partial f_{\rm GW}}{\partial q}
=0 \,,
\label{eq:Boltzmann}
\end{equation}
where $\Phi$ and $\Psi$ are the scalar gravitational potentials. The terms in square brackets describe the gravitational redshift experienced by gravitons during propagation.
Here, we consider only the positive-frequency component, $f=q/(2\pi)>0$. For this component, the spatial Fourier wave vector $\bq$ introduced in Sec.~\ref{sec:ffn relations} coincides with the comoving graviton momentum, so that $\hq$ is the propagation direction and $\hn=-\hq$. The negative-frequency component is obtained by $\hq\to-\hq$ in the direction-dependent terms. 

Analogous to the CMB, we decompose the distribution function as
\begin{equation}
f_{\rm GW}(\tau,\bx,\bq) = \bar{f}_{\rm GW}(q) - q\frac{\partial \bar{f}_{\rm GW}}{\partial q}\Gamma(\tau,\bx,\bq) + {\cal O}(\Phi^2) \,,
\label{eq:fdecomp}
\end{equation}
where the function $\Gamma$ parametrizes the first-order perturbation, which gives the leading term for the inhomogeneous graviton distribution. 
The Fourier transform of its spatial dependence is 
\begin{align}
\Gamma(\tau,\bx,\bq)
=
\int\frac{\rd^3\bk}{(2\pi)^3}\,
\Gamma(\tau,\bk,\bq)
e^{i\bk\cdot\bx} \,.
\end{align}
Here, $\bq$ labels the short GW wave vector, whereas $\bk$ is the wave vector of the spatial modulation of their distribution. As we can see from the perturbed Boltzmann equation \eqref{eq:Boltzmann}, the spatial dependence of $\Gamma$ is induced by the long-wavelength gravitational potentials, so that $\Gamma(\tau,\bk,\bq)$ is sourced by $\Phi(\tau,\bk)$ and $\Psi(\tau,\bk)$ with the same wave vector. Thus, in the CFF language, $\bq$ labels the short GW sector whose local statistics are measured, while $\bk$ is identified with the long-mode wave vector $\bk_L$ that modulates those statistics.

Integrating the Boltzmann equation yields
\begin{align}
\Gamma(\tau_0,\bk, \bq)=\int^{\tau_0}_{\tau_{\rm in}}d\tau_s \, e^{-i \bk \cdot \hq (\tau_0 - \tau_s)}\Bigg\{&\left[\Gamma(\tau_s,\bk,\bq)+\Phi(\tau_s,\bk)\right]\delta(\tau_s -\tau_{\text{in}}) + \Psi'(\tau_s,\bk)+\Phi'(\tau_s,\bk)
\Bigg\} \,,
\label{eq:Gamma_solution}
\end{align}
where $\tau_{\rm in}$ is the initial conformal time. 
Unlike the CMB line-of-sight solution, where the visibility function selects an effective emission epoch, the absence of GW scattering leaves the boundary contribution at $\tau_{\rm in}$ explicit. 
Determining the appropriate matching time and deriving its boundary condition from the preceding wave dynamics require separate consideration, 
particularly when GW production occurs near horizon crossing,
where the geometric-optics approximation is not manifestly valid. 
We leave this matching problem aside here, 
regard $\tau_{\rm in}$ and $\Gamma(\tau_{\rm in},\bk,\bq)$ as given, 
and focus below only on the propagation effects sourced by $\Phi$ and $\Psi$.

In the linear regime, the scalar potentials can be expressed in terms of the primordial curvature perturbation $\zeta(\bk)$ via transfer functions,
\begin{equation}
\Phi(\tau_s,\bk)=T_{\Phi}(k, \tau_s)\zeta(\bk)\,, \quad \Psi(\tau_s,\bk)=T_{\Psi}(k, \tau_s)\zeta(\bk) \,.
\label{eq:transfer_scalar}
\end{equation}
The SGWB anisotropies are related to $\Gamma$ by 
\begin{equation}
\delta^{\rm GW}(f,\hn)
= \left.\left[4-\frac{\partial \ln\bar{\Omega}_{\rm GW}(q)}{\partial\ln q}\right]\Gamma(\tau_0,\bx_0, q,\hq)\right|_{q=2\pi f,\,\hq=-\hn} \,,
\label{eq:delta_Gamma}
\end{equation}
where $\bar{\Omega}_{\text{GW}}(q)$ is the isotropic (background) energy density parameter \eqref{eq:GW energy density} and $\bx_0$ denotes the observer position. Throughout the Letter and the Supplemental Material, we adopt observer-centered coordinates, $\bx_0=\mathbf{0}$. This convention is implicit in the Fourier-space solution \eqref{eq:Gamma_solution} and the Letter.  
Then, the propagation-induced anisotropies are provided by 
\begin{align}
\delta^{\rm GW}(f,\hn)
= \left[4-\frac{\partial \ln\bar{\Omega}_{\rm GW}(q)}{\partial\ln q}\right]
\int \frac{d^3\bk}{(2\pi)^3} \int^{\tau_0}_{\tau_{\rm in}}d\tau_s\, e^{-i \bk \cdot \hq (\tau_0 - \tau_s)}
\left[T_\Phi(k, \tau_s)\delta(\tau_s-\tau_{\rm in}) + T'_\Phi(k,\tau_s) + T'_\Psi(k,\tau_s) \right]\zeta(\bk) \,,
\label{eq:delta_scalar}
\end{align}
with $q=2\pi f,\,\hq=-\hn$. 
Comparing Eq.~\eqref{eq:delta_scalar} with the CFF definition in the Letter, 
we can directly read off the CFF for the propagation-induced anisotropies as
\begin{equation}
F(f,\hn,\bk_L) =  
\left[4-\frac{\partial \ln\bar{\Omega}_{\rm GW}(f)}{\partial\ln f}\right]
\int^{\tau_0}_{\tau_{\rm in}}d\tau_s\, e^{i \hn \cdot \bk_L (\tau_0 - \tau_s)}
\left[T_\Phi(k_L, \tau_s)\delta(\tau_s-\tau_{\rm in}) + T'_\Phi(k_L,\tau_s) + T'_\Psi(k_L,\tau_s) \right] \,.
\label{eq:CFF_propagation}
\end{equation}
Thus, the CFF has the canonical form given in the Letter with 
\begin{equation}
\FFK(f,\tau_s) = \left[4-\frac{\partial \ln\bar{\Omega}_{\rm GW}(f)}{\partial\ln f}\right]\delta(\tau_s-\tau_{\rm in})\,, 
\quad \TF_\phi(k,\tau_s) = T_\Phi(k, \tau_s) \,,
\label{eq:SW CFF}
\end{equation}
for the SW effect, and
\begin{equation}
\FFK(f,\tau_s) = \left[4-\frac{\partial \ln\bar{\Omega}_{\text{GW}}(f)}{\partial\ln f}\right] \,,
\quad 
\TF_\phi(k, \tau_s) = T'_\Phi(k,\tau_s) + T'_\Psi(k,\tau_s) \,,
\label{eq:ISW CFF}
\end{equation}
for the ISW effect. 
Here, $\TF_\phi$ in Eq.~\eqref{eq:ISW CFF} denotes the effective transfer function from the primordial curvature perturbation to the ISW source $\Phi'+\Psi'$. The early-time ISW contribution generated before recombination falls within the universality class studied in the Letter: for the relevant large-angle modes, $k_L\tau_s\ll1$, and the transfer function is effectively scale independent. 
The late-time contribution, while outside the primordial regime targeted here, is not an independent unknown. In standard cosmology, it is sourced by the same gravitational potential that produces the ISW effect in the CMB, with the SGWB response differing only by the frequency-dependent factor. It can therefore, in principle, be reconstructed and modeled using cross-correlations with the CMB and low-redshift tracers, providing a template that can be subtracted or marginalized over in searches for primordial SGWB anisotropies. 

\subsection{Example II: Fossil effects in scalar-induced gravitational waves}

We now turn to the second mechanism for SGWB anisotropies, namely fossil-type effects. 
We illustrate the modulation of the source statistics of SIGWs in the presence of the primordial local-type scalar bispectrum. 
Our treatment primarily follows the works of Li {\it et al.}~\cite{Li:2023xtl, Li:2023qua} and Bartolo {\it et al.}~\cite{Bartolo:2019zvb}.

The evolution of SIGWs is governed by
\begin{equation}
    h''_\lambda + 2\mathcal{H}h'_\lambda+q^2 h_\lambda
    =4S_\lambda,
\label{eq:SIGW_EOM}
\end{equation}
where $\mathcal{H}$ is the conformal Hubble parameter. 
The source term $S_\lambda$ is given by
\begin{align}
    S_\lambda(\tau,\bq) &= \int\frac{\rd^3 \bq_a}{(2\pi)^3}e^\lambda_{ij}(\hq) q^i_a q^j_a\Bigg\{ 2\Phi(\tau,\bq-\bq_a)\Phi(\tau,\bq_a) \nonumber \\
    &+\frac{4}{3(1+\omega)\mathcal{H}^2}\left[\Phi'(\tau,\bq-\bq_a)+\mathcal{H}\Phi(\tau,\bq-\bq_a)\right]
    \left[\Phi'(\tau,\bq_a)+\mathcal{H}\Phi(\tau,\bq_a)\right]\Bigg\},
\label{eq:S_lambda}
\end{align}
where $\omega$ is the equation-of-state parameter of the background fluid. 
Here, we neglect anisotropic stress and set $\Psi = \Phi$ at linear order. 
We relate the scalar potential $\Phi$ to the primordial curvature perturbation $\zeta$ via the transfer function,
\begin{equation}
    \Phi(\tau, \bq)=T_{\Phi}(q, \tau)\zeta(\bq) \,.
\label{eq:transfer_scalar_fossil}
\end{equation}

Here, we suppose that the statistics of the curvature perturbations are non-Gaussian with the local-type bispectrum, which is parametrized as
\begin{equation}
    \zeta = \zeta_g
    + \frac{3}{5}f_{\rm NL}\zeta_g^2 \,,
\label{eq:zeta_local}
\end{equation}
in real space, where $\zeta_g$ satisfies the Gaussian statistics. 
Decomposing the field into short and long modes, $\zeta = \zeta_S + \zeta_L$, 
we can see that, in the presence of the non-Gaussianity \eqref{eq:zeta_local}, long modes $\zeta_L$ modulate the statistics of short modes $\zeta_S$, 
\begin{equation}
    \zeta_S = \zeta_{S, g}
    + \frac{3}{5}f_{\rm NL}\zeta_{S,g}^2 + \frac{6}{5}f_{\rm NL}\zeta_{L, g} \zeta_{S,g} = \zeta_{S, g}
    + \frac{3}{5}f_{\rm NL}\zeta_{S,g}^2 + \frac{6}{5}f_{\rm NL}\zeta_{L} \zeta_{S,g} + {\cal O}(f_{\rm NL}^2\zeta^3) \,,
\label{eq:zeta_short_long}
\end{equation}
After some algebra, the resulting angular anisotropy of the SIGW energy density at the source position $\bx_s$ can be expressed as~\cite{Bartolo:2019zvb}
\begin{equation}
      \delta^{\rm GW}(\tau_{\rm in}, \bx_s) = \frac{24}{5}f_{\rm NL}\int\frac{\rd^3 \bk}{(2\pi)^3}e^{i\bk \cdot \bx_s}\zeta_L(\bk) \,,
\label{eq:delta_fossil_simple}
\end{equation}
where $\tau_{\rm in}$ is the conformal time when the SIGW generation is terminated.  
The time $\tau_{\rm in}$ should be after which the geometrical optics approximation becomes valid, {\it i.e.}, when the GW modes have re-entered the Hubble horizon and the subsequent evolution can be treated as free propagation.
The result \eqref{eq:delta_fossil_simple} can also be interpreted from the response structure of the source term, 
    \begin{align}
        \frac{\delta S_\lambda(\tau, \bq)}{\delta \zeta_L(\bk)} = \frac{12}{5} f_{\rm NL} T_\Phi(k,\tau) \left. S_\lambda(\tau, \bq) \right|_{\zeta_S = \zeta_{S,g}} \,,
    \end{align}
where we have kept the leading-order term in the gradient expansion, $\zeta(\bq - \bk) \simeq \zeta(\bq)$. 
More detailed computation leads to~\cite{Li:2023qua}
\begin{equation}
    \delta^{\rm GW}(\tau_{\rm in}, \bx_s) = 
    \frac{3}{5}f_{\rm NL}\left[\frac{\Omega_{ng}(\tau_{\rm in},q)}{\bar{\Omega}_{\rm GW}(\tau_{\rm in},q)}\right]
    \int\frac{\rd^3 \bk}{(2\pi)^3}e^{i\bk \cdot \bx_s}\zeta_L(\bk) \,,
\label{eq:delta_fossil_general}
\end{equation}
where $\Omega_{ng}(\tau,q)$ encodes the response of the SIGW energy-density spectrum to the local non-Gaussian modulation, 
while $\bar{\Omega}_{\rm GW}(\tau,q)$ denotes the total isotropic SIGW energy-density spectrum. 
These results contribute to the boundary term in Eq.~\eqref{eq:Gamma_solution} (see also Eq.~\eqref{eq:delta_Gamma}). 
Thus, the observed SGWB anisotropy is provided by
    \begin{align}
        \delta^{\rm GW}(f, \hn) = \frac{3}{5}f_{\rm NL}\left[\frac{\Omega_{ng}(\tau_{\rm in},2\pi f)}{\bar{\Omega}_{\rm GW}(\tau_{\rm in},2\pi f)}\right]
        \int\frac{\rd^3 \bk}{(2\pi)^3}
        \int^{\tau_0}_{\tau_{\rm in}}d\tau_s \, e^{i \bk \cdot \hn (\tau_0 - \tau_s)} \delta(\tau_s-\tau_{\rm in}) \zeta_L(\bk) \,.
        \label{eq:delta_fossil}
    \end{align}

Comparing Eq.~\eqref{eq:delta_fossil} with the CFF definition in the Letter, 
we can directly read off the CFF for the fossil-type anisotropies as
\begin{equation}
    F(f,\hn, \bk_L) =
    \frac{3}{5}f_{\rm NL}\left[\frac{\Omega_{ng}(\tau_{\rm in},2\pi f)}{\bar{\Omega}_{\rm GW}(\tau_{\rm in},2\pi f)}\right]
        \int^{\tau_0}_{\tau_{\rm in}}d\tau_s \, e^{i \bk_L \cdot \hn (\tau_0 - \tau_s)} \delta(\tau_s-\tau_{\rm in}) \,.
\label{eq:CFF_fossil}
\end{equation}
Thus, the CFF has the canonical form given in the Letter with
\begin{equation}
    \FFK(f,\tau_s) = \frac{3}{5}f_{\rm NL}\frac{\Omega_{ng}(\tau_s,2\pi f)}{\bar{\Omega}_{\rm GW}(\tau_s,2\pi f)}\delta(\tau_s - \tau_{\rm in}) \,, \quad \TF_\phi(k, \tau_s) = 1 \,.
\label{eq:kernel_fossil}
\end{equation}
Here, we identify the primordial long mode with $\phi=\zeta$; its transfer function is unity for the superhorizon modes considered here.
This example thus confirms that the CFF formalism provides a unified description for fossil-type anisotropies as well, with the kernel now encoding the non-Gaussian modulation at the source epoch.

\end{document}